\documentclass[letterpaper,floatfix,aps,pra,amsmath,amssymb,twocolumn,
showpacs,footinbib]{revtex4-1}

\usepackage{verbatim}
\usepackage{color}
\usepackage{array}
\usepackage{amsmath}

\usepackage{graphicx}
\usepackage{hyperref}

\newcommand{\be}{\begin{equation}}
\newcommand{\ee}{\end{equation}}
\newcommand{\ba}{\begin{equation}}
\newcommand{\ea}{\end{equation}}
\newcommand{\bea}{\begin{eqnarray}}
\newcommand{\eea}{\end{eqnarray}}
\newcommand{\w}{\omega}

\newcommand{\td}{\tilde{d}}
\newcommand{\tu}{\tilde{U}}
\newcommand{\te}{\tilde{\epsilon}}
\newcommand{\tez}{\tilde{\epsilon}_{z,0}}

\newcommand{\tz}{\tilde{z}} 
\newcommand{\tze}{\tilde{\zeta}}

\newcommand{\bg}{\bold{G}}

\newcommand{\e}{\epsilon}
\newcommand{\Ai}{\mathrm{Ai}}
\newcommand{\Bi}{\mathrm{Bi}}

\newcommand{\Int}{\mathrm{int}}
\newcommand{\eref}[1]{Eq.~(\ref{#1})}
\newcommand{\esref}[1]{Eqs.~(\ref{#1})}

\newcommand{\ocite}[1]{Ref.~\cite{#1}}

\begin{document}

%\title{Magnetic and electrical control of valley %splitting in Si spin qubits
%with ideal and disordered interface}

\title{Electromagnetic control of valley splitting in ideal and disordered Si quantum dots}

\author{Amin Hosseinkhani}
\email{amin.hosseinkhani@uni-konstanz.de}
\affiliation{Department of Physics, University of Konstanz, D-78457 Konstanz, Germany}
\author{Guido Burkard}
\email{guido.burkard@uni-konstanz.de}
\affiliation{Department of Physics, University of Konstanz, D-78457 Konstanz, Germany}

\begin{abstract}
In silicon spin qubits, the valley splitting must be tuned far away from the qubit Zeeman splitting to prevent fast qubit relaxation.  In this work, we study in detail how the valley splitting depends on the electric and magnetic fields as well as the quantum dot geometry for both ideal and disordered Si/SiGe interfaces. We theoretically model a realistic electrostatically defined quantum dot and find the exact ground and excited states for the out-of-plane electron motion. This enables us to find the electron envelope function and its dependence on the electric and magnetic fields. For a quantum dot with an ideal interface, the slight cyclotron motion of electrons driven by an in-plane magnetic field slightly increases the valley splitting. Importantly, our modeling makes it possible to analyze the effect of arbitrary configurations of interface disorders. In agreement with previous studies, we show that interface steps can significantly reduce the valley splitting. Interestingly, depending on where the interface steps are located, the magnetic field can increase or further suppress the valley splitting.  Moreover, the valley splitting can scale linearly or, in the presence of interface steps, non-linearly with the electric field.
\end{abstract}

%\date{\today}
\maketitle

\section{Introduction}
\label{sec:intro}
The spin of isolated electrons trapped in silicon-based heterostructures is very promising for building high performance and scalable qubits \cite{Zwanenburg13}. The long relaxation time \cite{Morello10, Yang2013, Borjans19} and dephasing time \cite{Assali11, Tyryshkin12, Steger12} that are achieved in these qubits are due to the weak spin-orbit interaction and nuclear zero-spin isotopes. Strong coherent coupling between Si spin qubits and photons using superconducting resonators has been realized \cite{Mi18, Samkharadze18} while the fidelities demonstrated for single and two-qubit gates are steadily improving \cite{Veldhorst14, Yoneda18, Zajac18, Watson18, Dzurak19_tqg}. Having mentioned all these advantages, we note that the nature of the degenerate conduction band minima, known as valleys, in bulk silicon poses a significant challenge for the operation and scalability of silicon spin qubits. It can be shown that a combination of biaxial strain as well as the sharp interface potential lifts the valley degeneracy in Si heterostructures, and gives rise to two low-lying states \cite{Zwanenburg13}. In general, a qubit performs well only when the qubit energy splitting is well-separated from any other energy scale in the environment. In Si spin qubits, the spin couples to the valley degree of freedom due to interface-induced spin-orbit interaction \cite{Golub04,Veldhorst15R, Ferdous18}. If the valley splitting becomes equal to the qubit Zeeman splitting, a condition known as spin-valley hotspot, the valley-spin mixing for the qubit excited state reaches its maximum and gives rise to a very fast qubit relaxation via electron-phonon interaction \cite{Yang2013,Huang14}. It is, therefore, of crucial importance to understand how  the valley splitting behaves as a function of parameters of the system; namely, electric and magnetic fields, the quantum dot geometry and the roughness at the Si/barrier interface. 

In studying the valley splitting, a suitable starting point is the effective mass theory that can be used  to obtain the electron envelope function. This envelope function, in turn, depends on the above-mentioned system parameters, and the valley splitting can be deduced from it.  As we review in Section~\ref{sec:ExEn}, in the absence of a magnetic field, the Hamiltonian describing the full envelope function is separable. While the in-plane envelope function is trivially given by the harmonic-oscillator wave function (due to the in-plane parabolic confinement), to our knowledge, the (ground state of the) out-of-plane envelope function has only been studied and approximated via variational methods \cite{Friesen07, Koiller2011, Culcer2012, Tariq} or by setting the barrier potential to infinity \cite{Ruskov2018}. However, the assumptions involved in these methods render them less accurate for higher electric fields. In this paper, we model a realistic potential profile for a SiGe/Si/SiGe quantum dot by taking into account both Si/SiGe interfaces as well as an interface between SiGe and the insulating layer hosting the gate  electrodes. Within this model, we then find the \textit{exact} solution for the ground state as well as excited state envelope functions for the out-of-plane electron motion.

In the presence of an in-plane magnetic field, a cyclotron motion of electrons takes place which tends to increase the electron probability amplitude at the Si/SiGe interface \cite{Jock2018}. This effect can, in turn, modify and increase the valley splitting. The magnetic field couples in-plane to out-of-plane degrees of freedom and thus prevents us from finding the exact solution for the electron envelope function. Using the exact excited states for the out-of-plane envelope function, we find the full envelope function in the presence of a magnetic field by applying perturbation theory. We show that an in-plane magnetic field indeed slightly increases the valley splitting; up to a few Tesla, the valley splitting increases quadratically with the magnetic field. Besides this, for a quantum dot with an ideal interface, i.e, no miscuts and steps at the interface, we find that the dominant contribution to the valley splitting scales linearly with the electric field, $F_z$. 

During the experimental process of fabricating silicon heterostructures, the formation of steps and miscuts at the Si/SiGe seems to be inevitable \cite{Zandvliet93, Hollmann19}. It has been shown that the presence of interface steps can severely suppress the valley splitting \cite{Friesen06,Goswami06, Tariq}. Here we again use the exact excited states for the out-of-plane envelope function in order to perturbatively treat the effects of interface steps to the envelope function. We argue that our modeling is applicable to any configuration for the interface disorder. We first study how the interface steps suppress the valley splitting in the absence of a magnetic field. We show that the valley splitting of a disordered quantum dot can scale either sub-linearly, linearly or super-linearly with the electric field, depending on the steps' configuration. We then consider the effects of an in-plane magnetic field. While it has been speculated that the magnetic field can increase the valley splitting in the presence of interface steps \cite{Friesen06,Goswami06}, interestingly, we find that the magnetic field can both increase or further suppress the valley splitting depending on the locations of the steps. 

This paper is structured as follows: In Section \ref{sec:ExEn} we present our model and find the exact solution for the out-of-plane electron motion. In Section~\ref{sec:EnvBCle} we obtain the envelope function in the presence of an in-plane magnetic field for a quantum dot with an ideal Si/SiGe interface. In Section~\ref{sec:EnvBDis} we extend our model to include the interface disorders and derive the envelope function for a certain configuration of  steps. In Section~\ref{sec:Dis} we build on our findings for the envelope function to obtain and discuss the valley splitting; in Secs.~\ref{sec:VS_Fz} and \ref{sec:VS_B_clean}, we study how the valley splitting of an ideal quantum dot depends on the electric and magnetic field field. In \ref{sec:VS_B_dis}, we consider interface disorder and calculate the valley splitting and its phase depending on the location of the step. We then investigate the role of the electric and magnetic fields in modifying the valley splitting. In Section~\ref{sec:conc} we summarize and conclude the paper. The Appendices contain further details of our analysis.

\section{Model}
\label{sec:model}
\begin{figure}[t!]%============================================
\begin{center}
\includegraphics[width=0.48\textwidth]{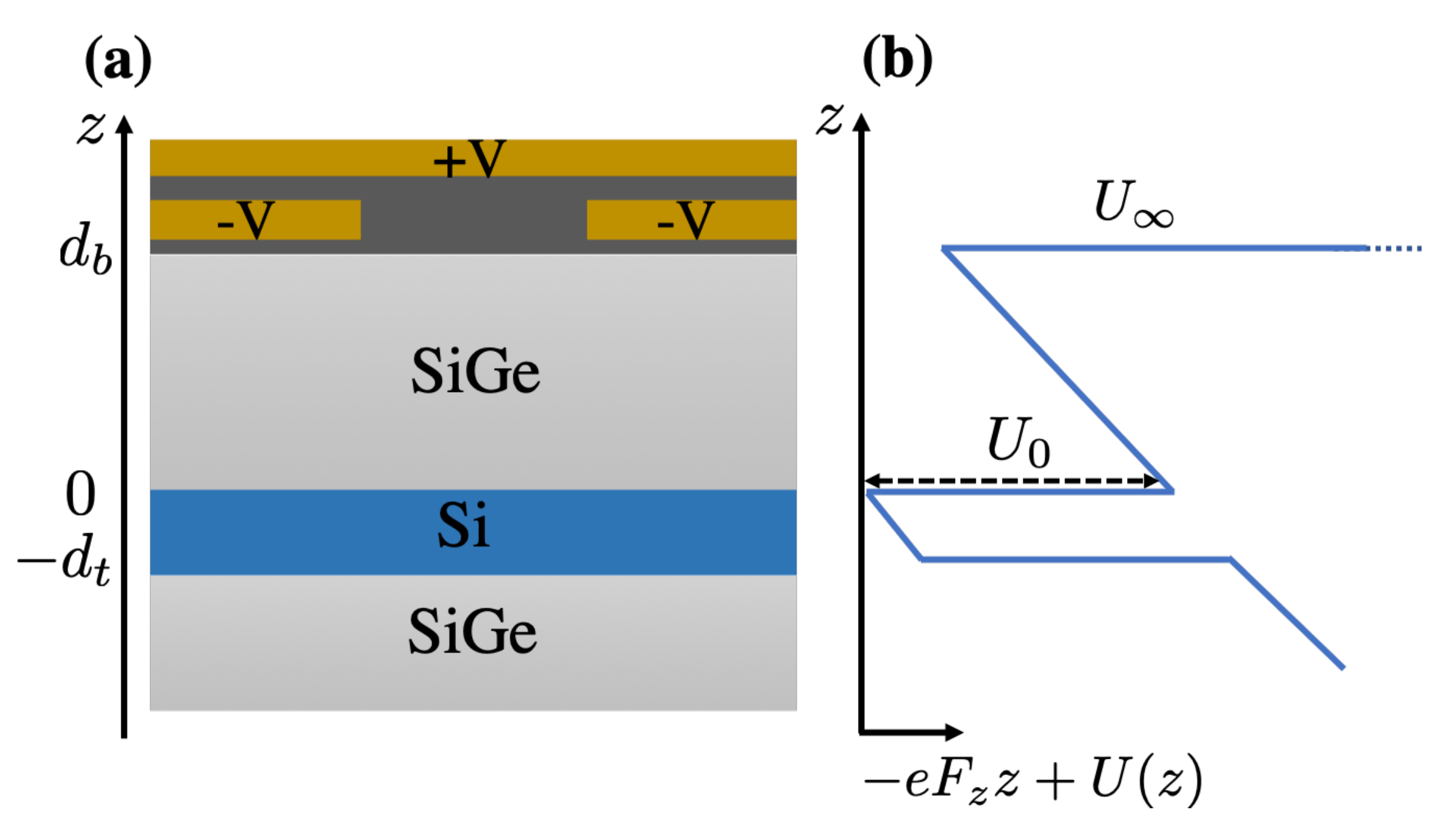}
\end{center}
\caption{\textbf{(a)} Schematic layered structure of a single SiGe/Si/SiGe quantum dot. The dark gray area is an insulating layer hosting the electric gates. The $\pm V$ gates are used to trap and confine a single electron in the silicon layer. \textbf{(b)} The electrostatic potential profile along the growth direction $\hat{z}$. $F_z$ is the out-of-plane electric field generated by the gates, $U_0$ is the potential barrier of the SiGe layers, $U_{\infty}$ is the infinite potential barrier due to the insulating layer. The potential energy $U(z)$ is defined in \eref{eq:Uz}.}  \label{fig:qubit_layout}
\end{figure}

We consider a SiGe/Si/SiGe heterostructure grown along the $\hat{z}$ direction $([001])$ while the silicon layer is between $-d_t<z<0$. Panel (a) of Figure~\ref{fig:qubit_layout} shows a schematic cross-section of the layer structure of the system. An electric field is applied along $\hat{z}$ via the gates, and we consider both interfaces between Si and SiGe at $z=0$ as well as $z=-d_t$. The energy offset between the minima of the conduction band in Si and SiGe is given by $U_0=150$ meV. Moreover, we also consider the interface between the top SiGe barrier and the insulating layer that hosts the electric gates. Inside the insulating layer, we take $U_\infty =\infty$ which indicates the envelope function does not leak into that region. Panel (b) of Figure~\ref{fig:qubit_layout} shows the full potential along $\hat{z}$. 
Moreover, here we assume a generally elliptical quantum dot with harmonic in-plane confinement. We denote the radius of the quantum dot along $\hat{x}$ by $x_0$ and the radius along $\hat{y}$ by $y_0$.
\subsection{Exact envelope function in absence of a magnetic field}
 %======================================
\label{sec:ExEn}
Within the effective mass theory and in the absence of a magnetic field, the Hamiltonian describing the envelope function reads,
\begin{align}
\label{eq:H_xyz}
H_{xyz}=&\frac{p_x^2}{2m_t}+\frac{1}{2}m_t\w_x^2x^2 +\frac{p_y^2}{2m_t}+\frac{1}{2}m_t\w_y^2y^2\nonumber\\&+\frac{p_z^2}{2m_l}-eF_zz+ U(z).
\end{align}
Here $m_t=0.19\:m_e$ and $m_l=0.98\:m_e$ are the transverse and longitudinal effective mass, $\w_{x}=2\hbar/m_tx_0^2$ and $\w_{y}=2\hbar/m_ty_0^2$ are the confinement frequencies along $\hat{x}$ and $\hat{y}$, and 
\begin{align}
\label{eq:Uz}
U(z)=U_0\theta(-z-d_t) + U_0\theta(z) + U_{\infty}\theta(z-d_b).
\end{align}

\eref{eq:H_xyz} clearly gives rise to a separable envelope function $\psi_{xyz}=\psi_x\psi_y\psi_z$ where $\psi_x$ and $\psi_y$ are the well-known harmonic oscillator wavefunctions. Our main objective in this section is to find the exact eigenstates $\psi_{z,n}$ and eigenenergies $E_{z,n}$ for the out-of-plane electron motion.

Given \eref{eq:H_xyz},  we write the Schr\"odinger equation for the envelope function $\psi_{z,n}$ as
\begin{align}
\label{eq:H_z}
\left\{\frac{p_z^2}{2m_l}-eF_zz+ U(z)\right\}\psi_{z,n}=E_{z,n}\psi_{z,n}.
\end{align}
We now use the electrical confinement length,
\begin{align}
\label{eq:z0}
z_0=\left[\frac{\hbar^2}{2m_leF_z}\right]^{1/3},
\end{align}
and its associated energy scale,
\begin{align}
\label{eq:e0}
\epsilon_0 = \frac{\hbar^2}{2m_lz_0^2},
\end{align}
in order to piecewise expressing \eref{eq:H_z} as,
\begin{align}\label{eq:zeta}
\frac{d^2}{d\tz^2}\psi_{z,n} &- \left(\tilde{U}_0-\tz-\te_{z,n}\right)\psi_{z,n}=0 , \: 0 < z < d_b, \nonumber\\
\frac{d^2}{d\tz^2}\psi_{z,n} &- \left(-\tz-\te_{z,n}\right)\psi_{z,n}=0 , \quad  -d_t\leq z \leq  0, \nonumber \\
\frac{d^2}{d\tz^2}\psi_{z,n} &- \left(\tilde{U}_0-\tz-\te_{z,n}\right)\psi_{z,n}=0, \:  z < -d_t\, .
\end{align}
Here $\tz=z/z_0$, $\te_{z,n}=E_{z,n}/\epsilon_0$ and $\tilde{U}_0=U_0/\epsilon_0$ are the normalized length, eigenenergy, and potential. 

The above equation, at each interval, has generally two linearly-independent solutions known as Airy functions of the first and second kind, Ai and Bi \cite{Davies}. We thus find the exact solution for $\psi_{z,n}$:
\be\label{eq:OP_step}
  \psi_{z,n} =\\
   N_0z_0^{-1/2}
  \begin{cases}
   c_1\Ai(\tze_{n}) + c_2 \Bi(\tze_{n})\:, &0 < z < d_b\, \\
     c_3 \Ai(\tze_{n})+ c_4 \Bi(\tze_{n})\:, & -d_t \leq z \leq  0 \, \\
     c_5 \Ai(\tze_{n})\:,  &z < -d_t\,
  \end{cases}
\ee
where we defined,
\be\label{eq:zeta}
  \tze_n=\\
  \begin{cases}
\tilde{U}_0-\tz-\te_{z,n}\: , & 0 < z < d_b\, \\
-\tz-\te_{z,n}\:                 , & -d_t\leq z \leq  0 \, \\
\tilde{U}_0-\tz-\te_{z,n}\:. &  z < -d_t .
  \end{cases}
\ee 
Note that the Bi function is omitted from the solution for $z < -d_t$. This is based on the physical ground that Bi does not give rise to a decaying behaviour inside the extended  barrier layer.

In order to find the eigenenergies and determine the coefficients involved in the envelope function \eref{eq:OP_step}, we note that $\psi_{z,n}$ and its first derivative must be continuous at the boundaries between Si and SiGe, i.e. at $z=-d_t$ and $z=0$.  Moreover, since there is no leakage to the insulating layer, the envelope function must vanish at $z=d_b$. By imposing these boundary conditions, we obtain the  equation below from which we can numerically find all possible eigenenergies,
\begin{align}
\label{eq:Eq_Ez}
f_1(\te_{z,n},\td_t,\tu_0)&g_1(\te_{z,n},\td_b,\tu_0)\nonumber\\ &-f_2(\te_{z,n},\td_t,\tu_0)g_2(\te_{z,n},\td_b,\tu_0)=0,
\end{align}
with the definitions,
\begin{align}
f_1(\te_{z,n},\td_t,\tu_0) =& \Bi'(\td_t-\te_{z,n})\Ai(\tu_0+\td_t-\te_{z,n})\nonumber\\
&+\Bi(\td_t-\te_{z,n})\Ai'(\tu_0+\td_t-\te_{z,n}),\\
f_2(\te_{z,n},\td_t,\tu_0) =& \Ai(\td_t-\te_{z,n})\Ai'(\tu_0+\td_t-\te_{z,n})\nonumber\\
&+\Ai'(\td_t-\te_{z,n})\Ai(\tu_0+\td_t-\te_{z,n}),
\end{align}
and,
\begin{align}
&g_1(\te_{z,n},\td_b,\tu_0) =\nonumber \\ & \Ai(-\te_{z,n})\left[\Bi'(\tu_0-\te_{z,n})-\Ai'(\tu_0-\te_{z,n})\frac{\Bi(\chi_{n,d_b})}{\Ai(\chi_{n,d_b})}\right]\nonumber \\
&-\Ai'(-\te_{z,n})\left[\Bi(\tu_0-\te_{z,n})-\Ai(\tu_0-\te_{z,n})\frac{\Bi(\chi_{n,d_b})}{\Ai(\chi_{n,d_b})}\right],\\
&g_2(\te_{z,n},\td_b,\tu_0) =\nonumber \\ & \Bi'(-\te_{z,n})\left[\Bi(\tu_0-\te_{z,n})-\Ai(\tu_0-\te_{z,n})\frac{\Bi(\chi_{n,d_b})}{\Ai(\chi_{n,d_b})}\right]\nonumber \\
&-\Bi(-\te_{z,n})\left[\Bi'(\tu_0-\te_{z,n})-\Ai'(\tu_0-\te_{z,n})\frac{\Bi(\chi_{n,d_b})}{\Ai(\chi_{n,d_b})}\right],
\end{align}
where $\chi_{n,d_b}=\tu_0-\td_b-\te_{z,n}$ and $\Ai '$ and $\Bi '$ are the first derivatives of the $\Ai$ and $\Bi$ functions.\\
\begin{figure}[t!]%============================================
\begin{center}
\includegraphics[width=0.48\textwidth]{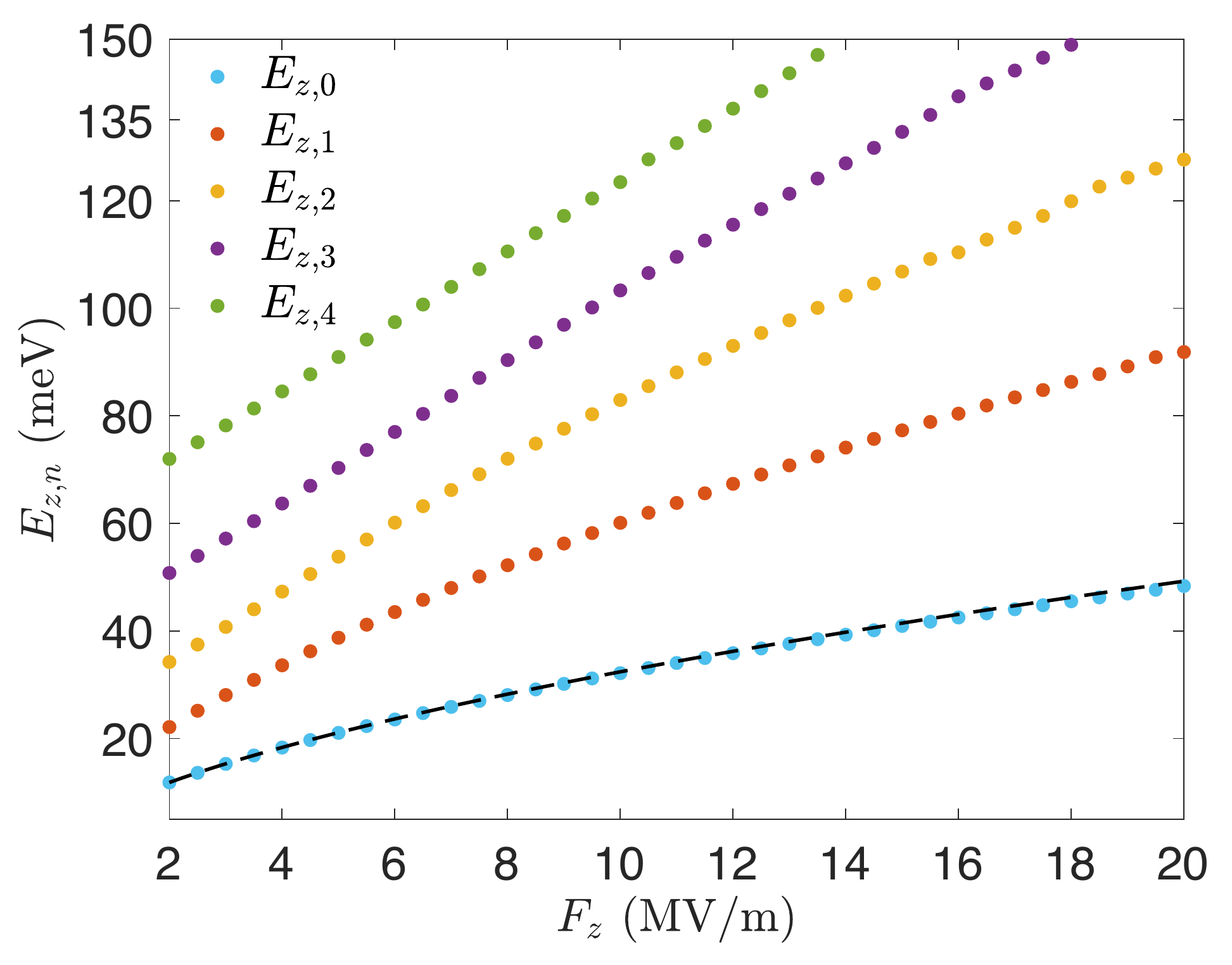}
\end{center}
\caption{Eigenenergies of the ground state $E_{z,0}$ and first few excited states up to the 4'th excited state $E_{z,4}$ as a function of the electric field $F_z$. The dots are obtain from numerically solving \eref{eq:Eq_Ez}.  The dashed line is the ground state energy obtained from \eref{eq:E_0z}. The used quantum dot parameters are $d_t=10$ nm and $d_b=46$ nm.}  \label{fig:E_z_vs_Fz}
\end{figure} 
%======================================----------
%======================================---------- 
\begin{figure}[t!]
\begin{center}

 \includegraphics[width=0.48\textwidth]{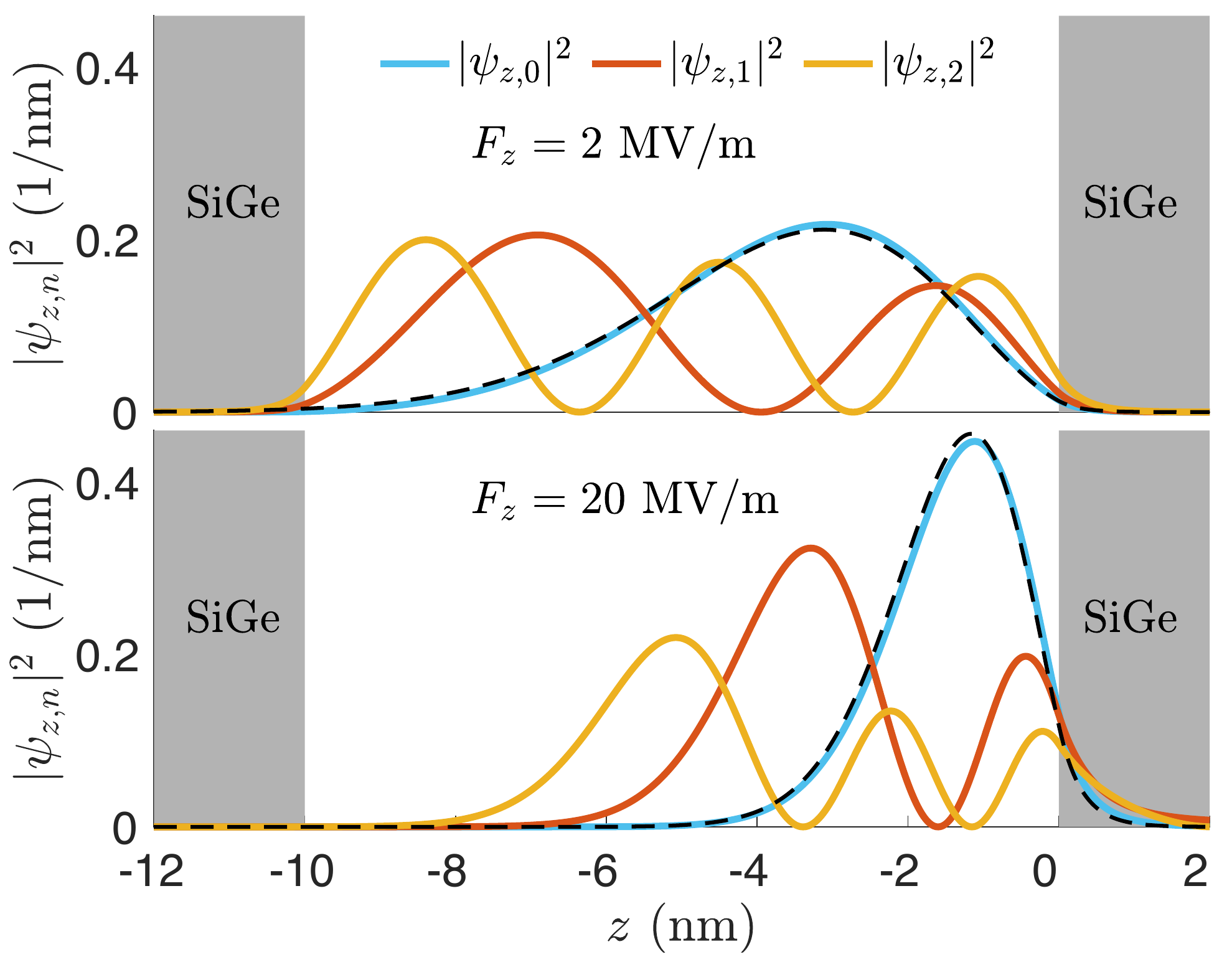}
\end{center}
\caption{Probability density of the ground state and the first and second excited states along $\hat{z}$. The gray areas mark the SiGe barriers. The solid lines show the exact envelope functions obtained from \eref{eq:OP_step} while the eigeneregies are numerically found from \eref{eq:Eq_Ez}; see Fig.~\ref{fig:E_z_vs_Fz}. The dashed lines show the approximate ground state envelope function given by \eref{eq:psi_0z}. As expected, the envelope functions are pushed upwards by increasing the electric field $F_z$. The parameters for the quantum dot are the same as noted in the caption of 
Fig.~\ref{fig:E_z_vs_Fz}.}  \label{fig:Psi_z_vs_Fz}
\end{figure} %======================================

Once the (normalized) eigenenergy $\te_{z,n}$ is found, we use it to calculate the coefficients $c_1$ to $c_5$. The coefficient $N_0$ is found by using the normalization of the envelope function.  We note that by solving \eref{eq:Eq_Ez}, we also find a set of states where the envelope function is  \textit{not} localized in the Si quantum well but rather in the upper SiGe barrier underneath the insulating layer. As we discuss it in Section \ref{sec:Dis}, the valley splitting is basically determined by the ground state localized in the Si quantum well. In the presence of a magnetic field or interface steps, we also need to take into account the excited states which have sizable overlap with the localized ground state in the Si quantum well; see \esref{eq:alpha_n}, (\ref{eq:beta_n}) and (\ref{eq:gamma_mn}). As such, the states that are localized underneath the insulating layer do not contribute to the behavior of the valley splitting, and we neglect them in this paper.

For the ground state of the electron motion along $\hat{z}$, we can simplify the analysis presented above and find analytic relations.  As we  show in Appendix \ref{app:GS_SC}, the (normalized) ground state energy in the regime of a deep quantum well, $\tu_0\gg 1$, can be expressed up to the leading order as
\begin{align}
\label{eq:E_0z}
\te_{z,0} = r_0 -\tilde{U}_0^{-1/2} + \mathcal{O}(\tilde{U}_0^{-3/2})
\end{align} 
where $-r_0\simeq-2.338$ is the smallest root (in absolute value) of the Ai function. The normalized envelope function in this case is approximated by,
\be\label{eq:psi_0z}
  \psi_{z,0}(\tz) \simeq\\
   \frac{z_0^{-1/2}}{\Ai'(-r_0)} % \frac{\Ai'(-\te_{z,0})}{\Ai(-\te_{z,0})}  (\Ai'(-\te_{z,0})/\Ai(-\te_{z,0}))
  \begin{cases}
      \Ai(-\te_{z,0})e^{-\frac{\Ai'(-\te_{z,0})}{\Ai(-\te_{z,0})}\tz}\: , & \tz > 0 \, \\
   \Ai(-\tz-\te_{z,0})\: , & \tz \leq 0\, 
  \end{cases}
\ee

In Figure~\ref{fig:E_z_vs_Fz} we show the obtained energies for the ground state as well as first few excited states as a function of the applied electric field.  In Figure~\ref{fig:Psi_z_vs_Fz} we also show the probability density $|\psi_{z,n}|^2$ for the ground state and first two excited states for two different electric fields. In both figures, a comparison between the numerics and the analytical relations Eqs.~(\ref{eq:E_0z}) and (\ref{eq:psi_0z}) for the ground state shows a very good agreement.  

We note here that the interface-induced spin-orbit interaction is neglected in our model. Consideration of this effect has been shown to be essential for explaining the valley-dependent \textit{g}-factor in silicon quantum dots \cite{Ferdous18,Ruskov2018}. However, as noted in \ocite{Ruskov2018}, the matrix elements involved in the spin-orbit interaction are much smaller than the valley splitting matrix element. This justifies our omission of the interface-induced spin-orbit interaction. As we will show in the next sections, the information stored in the excited states $\psi_{z,n\geq 1}$ enables us to obtain the full envelope function $\psi_{xyz,0}$ in a finite magnetic field, and also makes it possible to study realistic cases where there are steps and miscuts at the Si/SiGe interface. 

\subsection{Envelope function in the presence of an in-plane magnetic field with ideal Si/SiGe interface}
\label{sec:EnvBCle}
Let us now consider a quantum dot with an ideally flat Si/SiGe interface in the presence of an in-plane magnetic field $\bold{B}=(B_x,B_y,0)$. We use a gauge for which the vector potential becomes $\bold{A}=(0,0,yB_x-xB_y)$. By substituting $p_z\rightarrow p_z-eA_z(\textbf{B})$ in \eref{eq:H_xyz}, we arrive at the following form for the Hamiltonian describing the envelope function,
\begin{align} 
\label{eq:Htbp}
H= H_0(\bold{B}) + H_{\mathrm{pert}}(\bold{B}).
\end{align}
where we start from the separable Hamiltonian,
\begin{align}
\label{eq:H_0_bp}
H_0(\bold{B})= &\frac{p_x^2}{2m_t} +\frac{1}{2}m_t\omega_{x}'^2(B_y)x^2 +  \frac{p_y^2}{2m_t}+\frac{1}{2}m_t\omega_{y}'^2(B_x)y^2\nonumber\\&+\frac{p_z^2}{2m_l}-eF_zz+U(z)\:,
\end{align}
and treat the field-induced couplings as a perturbation,
\begin{align}
\label{eq:H_1_bp}
H_{\mathrm{pert}}(\bold{B})= &-B_x\frac{e}{m_l}yp_z+B_y\frac{e}{m_l}xp_z -B_xB_y\frac{e^2}{m_l}xy.
\end{align}
We note that the confinement frequencies and lengths along $\hat{x}$ and $\hat{y}$ are modified by the magnetic field,
\begin{align}
&\omega_{x}'(B_y)=\omega_{x}\left(1+\frac{e^2B_y^2}{m_tm_l\omega_{x}^2}\right)^{1/2},\\
&\omega_{y}'(B_x)=\omega_{y}\left(1+\frac{e^2B_x^2}{m_tm_l\omega_{y}^2}\right)^{1/2},\\\label{eq:x0'}
&x_0'(B_y)=x_0\left(1+\frac{e^2B_y^2}{4\hbar^2}\frac{m_t}{m_l}x_0^4\right)^{-1/4},\\\label{eq:y0'}
&y_0'(B_x)=y_0\left(1+\frac{e^2B_x^2}{4\hbar^2}\frac{m_t}{m_l}y_0^4\right)^{-1/4}.
\end{align} 

\begin{figure}[t!]
\begin{center}
\includegraphics[width=0.48\textwidth]{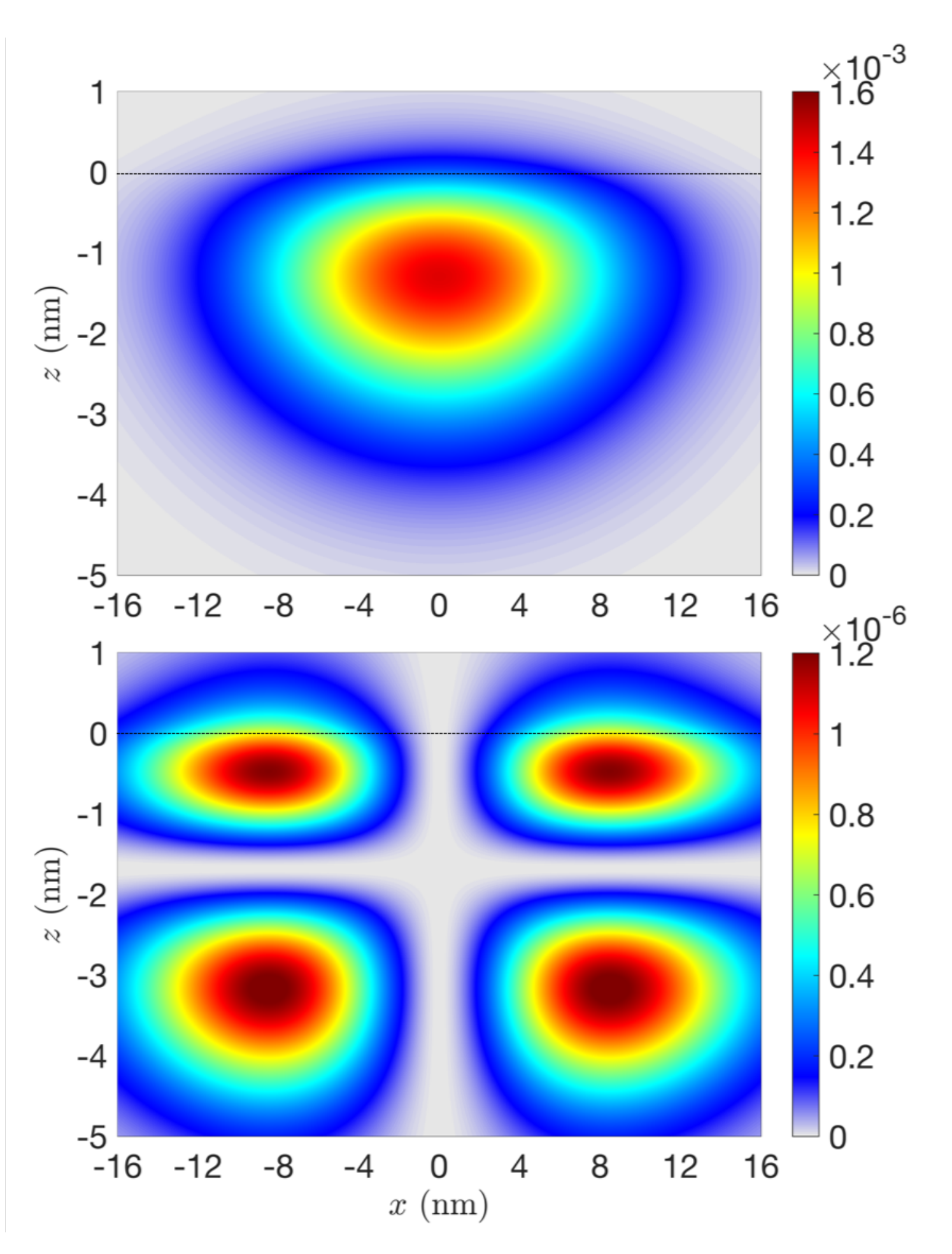}
\end{center}
\caption{Upper panel: The electron probability density of a quantum dot with an ideal interface in the $x-z$ plane in leading order, $|\psi_{xyz,0}^{(\bold{0})}(\bold{B})(x,y=0,z)|^2$ (1/nm$^3$); see \eref{eq:Psi_B_0}. Lower panel: The correction to the probability amplitude in the $x-z$ plane, $|\psi_{xyz,0}^{(\bold{1})}(\bold{B})(x,y=0,z)|^2$ (1/nm$^3$)
due to an in-plane magnetic field; see  \eref{eq:Psi_B_1}. The parameters used are the same as given in the caption of Table~\ref{tab:param}. The dashed line in both panels marks the ideally flat Si/SiGe interface. }  \label{fig:2D_psi}
\end{figure}

In order to obtain the envelope function from \eref{eq:Htbp}, we treat $H_0(\bold{B})$ exactly and apply perturbation theory in $H_{\mathrm{pert}}(\bold{B})$. The ground state up to the first order perturbation then reads,
\begin{align}
\label{eq:Psi_B}
\Psi_{xyz,0}(\bold{B})=\psi_{xyz,0}^{(\bold{0})}(\bold{B})+\psi_{xyz,0}^{(\bold{1})}(\bold{B}),
\end{align}
where,
\begin{align}\label{eq:Psi_B_0}
\psi_{xyz,0}^{(\bold{0})}(\bold{B})=&\psi_{x,0}(B_y)\psi_{y,0}(B_x)\psi_{z,0}\:,
\end{align}
and,
\begin{align}\label{eq:Psi_B_1}
\psi_{xyz,0}^{(\bold{1})}(\bold{B})=
&-iB_x\frac{y_0'}{z_0}\psi_{x,0}(B_y)\psi_{y,1}(B_x)\sum_{n=1}^{n_{\rm max}}\alpha_n\psi_{z,n}\nonumber\\
&+iB_y\frac{x_0'}{z_0}\psi_{x,1}(B_y)\psi_{y,0}(B_x)\sum_{n=1}^{n_{\rm max}}\beta_n\psi_{z,n}\nonumber\\
&-B_xB_yx_0'y_0'\eta\psi_{x,1}(B_y)\psi_{y,1}(B_x)\psi_{z,0}\:,
\end{align}
where the number of relevant bound  excited states
in the vertical direction for $F_z=15\,{\rm MV/m}$ is found to be ${n_{\rm max}}=3$,
see Fig.~\ref{fig:E_z_vs_Fz}.
Here we defined the coefficients,
\begin{align}
\label{eq:alpha_n}
\alpha_n&=-\frac{1}{2}\hbar\frac{e}{m_l}\frac{\langle\psi_{z,0}|\partial/\partial \tz|\psi_{z,n}\rangle}{E_{z,0}-E_{z,n}-\hbar\w_y'},\\ \label{eq:beta_n}
\beta_n&=-\frac{1}{2}\hbar\frac{e}{m_l}\frac{\langle\psi_{z,0}|\partial/\partial \tz|\psi_{z,n}\rangle}{E_{z,0}-E_{z,n}-\hbar\w_x'},\\
\eta&=-\frac{1}{4}\frac{e^2}{m_l}\frac{1}{\hbar\w_x'+\hbar\w_y'}.
\end{align}
We numerically calculate $\alpha_n$ and $\beta_n$ using the excited states $\psi_{z,n}$ obtained in Section~\ref{sec:ExEn}. For a circular dot we obtain $\alpha_n=\beta_n$. For an elliptical dot with realistic parameters, these coefficients remain close to each other since the confinement along $\hat{z}$ in quantum dots is always stronger than the in-plane confinements.  Table~\ref{tab:param} shows an example for the values of  $\alpha_n$ and $\beta_n$. With the set of parameters used in Table~\ref{tab:param}, we find $x_0'y_0'\eta=-8.64\times10^{-4}\mathrm{T}^{-2}$. Therefore, $(y'_0/z_0)\alpha_1$ and $(x'_0/z_0)\beta_1$ in \eref{eq:Psi_B_1} are one order of magnitude larger than $x_0'y_0'\eta$.
In Figure~\ref{fig:2D_psi} we  show the probability density in the $x-z$ plane in leading order, $|\psi_{xyz,0}^{(\bold{0})}(\bold{B})|^2$, as well as the first order correction, $|\psi_{xyz,0}^{(\bold{1})}(\bold{B})|^2$.
%============================================
\begin{table}[t!]
\begin{center}
\begin{tabular}{|c||c c|} \hline
$n$ & $\alpha_n\: (10^{-4}\mathrm{T}^{-1})$ & $\beta_n\: (10^{-4}\mathrm{T}^{-1})$ \\ \hline \hline
1 & -9.18 & -7.15 \\ \hline
2 & 2.85 &  2.26 \\ \hline
3& -1.38 & 	-1.10\\ \hline
\end{tabular}
\caption{\label{tab:param} The coefficients $\alpha_n$ and $\beta_n$ (in units of inverse Tesla). Here we used $F_z=15$ MV/m (corresponding to $z_0=1.40$ nm), $x_0=12$ nm, $y_0=15$ nm, and $B_x=B_y=5$ T.}
\end{center}
\end{table}
%============================================
%============================================
 %======================================

\subsection{Envelope function with disordered Si/SiGe interface}
\label{sec:EnvBDis}
So far, we have studied structures where the interface between Si and SiGe is perfectly flat and is located at $z=-d_t$ and $z=0$. However, during the experimental fabrication of Si qubit nanostructures, the formation of miscuts and steps at the interfaces is highly probable. Such uncontrolled disorder can modify the valley splitting and its phase, and is considered to be the main reason that makes the valley splitting a device-dependent quantity. In \ocite{Tariq}, several configurations for the steps at the interface are considered, and in each case, the envelope function is formed from a variational ansatz that uses a smooth interpolation between the envelope functions far from the step position (i.e. envelope functions for perfect interface).

In this section, we extend our model to include stair-like interface steps parallel to the $\hat{y}$ axis at the upper Si/SiGe interface as depicted in Figure~\ref{fig:layer_disor}. Our main objective here is to study how these miscuts  influence the quantum dot envelope wavefunction. Note that  disorder could also be present at the lower SiGe/Si interface. However, since the amplitude of the envelope function is small at the lower interface, the effects of possible disorder is negligible.  We also point out that other disorder configurations at the upper Si/SiGe interface can  be analyzed using the same approach.

In silicon, the thickness of each atomic layer is $a_0/4$ where $a_0=0.543$ nm denotes the lattice constant. This indicates that the change in the interface position due to a few miscuts is much smaller than the total thickness of the envelope function along $\hat{z}$ and enables us to use perturbation theory in order to obtain the electron envelope function. We take the $z$-position of the interface layer that contains the quantum dot center as the reference for the position of the barrier interface (e.g., the layer within $[x_2,x_3]$ in Fig.~\ref{fig:layer_disor}), and take any change to the interface position due to the miscuts as a perturbation. We describe the disordered SiGe/Si/SiGe interface  with the step potential
\begin{align}
\label{eq:Udis}
U_{\mathrm{dis}}(x,z)=U_0\theta(-z-d_t)+ U_0\theta(z)+U_{\mathrm{steps}}(x,z),
\end{align}
where,
\begin{align}
\label{eq:U_pert}
U_{\mathrm{steps}}(x,z)=U_0\Big[&\theta(-z)\theta(z+\frac{a_0}{4})\theta(x-x_1)\theta(x_2-x) \nonumber\\
  +&\theta(-z)\theta(z+\frac{a_0}{2})\theta(x_1-x) \nonumber\\
  -&\theta(z)\theta(z-\frac{a_0}{4})\theta(x-x_3)\theta(x_4-x) \nonumber\\
  -&\theta(z)\theta(z-\frac{a_0}{2})\theta(x-x_4) \Big].
\end{align}

The Hamiltonian describing the envelope function with disordered interface at finite in-plane magnetic field can again be written in the form of \eref{eq:Htbp} where, in this case, \eref{eq:U_pert} is added to the perturbative part of the Hamiltonian, \eref{eq:H_1_bp}. The ground-state envelope function then reads up to the \textit{second}-order perturbation with respect to the interface disorders,
%============================================
\begin{figure}[t!]
\begin{center}
\includegraphics[width=0.35\textwidth]{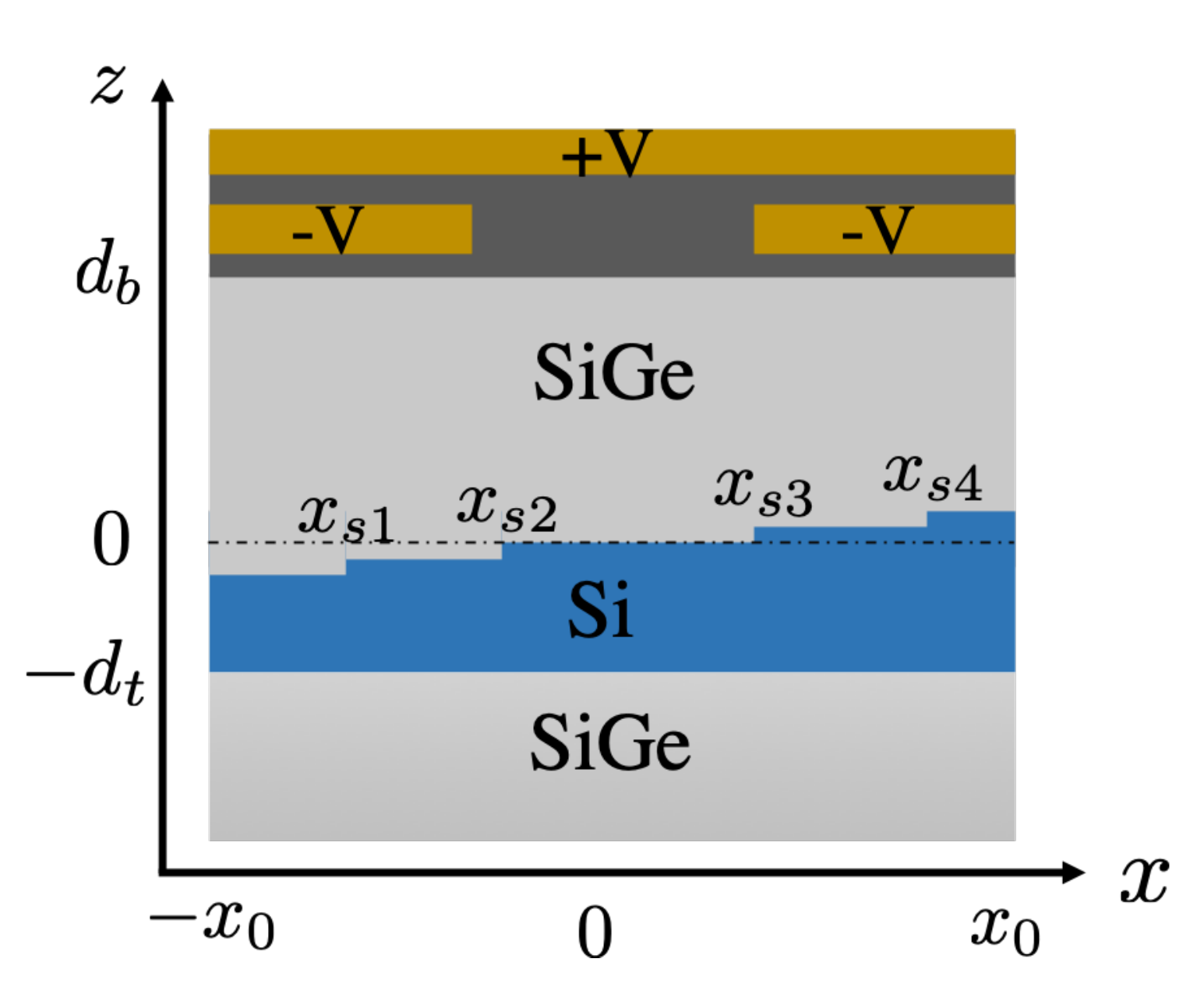}
\end{center}
\caption{Schematic layered structure of a quantum dot with stair-like disordered Si/SiGe interface.}  \label{fig:layer_disor}
\end{figure} %======================================
\begin{align}\label{eq:Psi_dis_B}
\Psi_{xyz,0}^{\mathrm{dis}}(\bold{B})=N_0\Big[&\psi_{xyz,0}^{(\bold{0})}(\bold{B})+\psi_{xyz,0}^{(\bold{1})}(\bold{B})+\mathcal{D}_{xyz,0}^{(\bold{1})}(\bold{B})\nonumber\\
&+\mathcal{D}_{xyz,0}^{(\bold{2})}(\bold{B})\Big]
\end{align}
where $N_0$ is a normalization constant and $\mathcal{D}_{xyz,0}^{(\bold{1})}$ is the first-order correction due to the interface disorder that amounts to
\begin{align}\label{eq:D_dis_1}
\mathcal{D}_{xyz,0}^{(\bold{1})}(\bold{B})=\psi_{y,0}(B_x)\sum_{\{m,n\}\neq \{0,0\}}\gamma_{m,n}\psi_{x,m}(B_y)\psi_{z,n},
\end{align}
for which the coefficients 
\begin{align}
\label{eq:gamma_mn}
\gamma_{m,n}=\frac{\langle\psi_{x,m}(B_y)\psi_{z,n}|U_{\mathrm{steps}}|\psi_{x,0}(B_y)\psi_{z,0}\rangle}{E_{0,z}-E_{n,z}-m\hbar\w_x'},
\end{align}
shall be calculated numerically. Table~\ref{tab:param_gamma} shows examples for the values of $\gamma_{m,n}$. Since the out-of-plane confinement is much stronger than the in-plane confinement, the largest contribution comes from $m=1$ and $n=0$. Moreover, we observe that by taking up to 4 excited states $\psi_{x,m}$, the values of $\gamma_{m,n}$ substantially decay.  As such, we can set $m_\mathrm{max}=4$ as a cutoff in the summation in \eref{eq:D_dis_1}.  
%============================================
\begin{table}[t!]
\begin{center}
\begin{tabular}{|c||c c c c|} \hline
$m$ & $\gamma_{m,0}$ & $\gamma_{m,1}$ & $\gamma_{m,2}$ & $\gamma_{m,3}$ \\ \hline \hline
0 & N/A          &   0.0170   &  -0.0082   &  0.0047      \\ \hline
1 &   0.4204   & -0.0564   &   0.0319    &  -0.0215      \\ \hline
2 &  -0.0353  & 0.0073     & -0.0036    &    0.0019      \\ \hline
3 &  -0.0214  & 0.0069     & -0.0043    &    0.0031      \\ \hline
4 &   0.0088   & -0.0029   &   0.0015    &  -0.0008      \\ \hline
5 & -0.0001   & -0.0001   &   0.0001    &  -0.0001      \\ \hline
6 &  -0.0025  & 0.0010     &  -0.0006   &  0.0003       \\\hline
7 &   0.0025   & -0.0012   &   0.0008    &  -0.0006      \\ \hline
8 &   0.0004   &-0.0002    &   0.0001    &   0.0000      \\ \hline
\end{tabular}
\caption{\label{tab:param_gamma}The coefficients $\gamma_{m,n}$. Here we assumed $x_{s1}=-7$ nm, $x_{s2}=-2$ nm, $x_{s3}=3$ nm, and $x_{s4}=7$ nm. The other parameters are the same as given in the caption of Table~\ref{tab:param}. We also find $\gamma_{1,0}'=0.3784$ and $\gamma_{2,0}'=0.0663$; see \eref{eq:gamma'}. }
\end{center}
\end{table}
%============================================

For the second-order correction due to the interface disorder, we only keep the leading-order terms to arrive at (see Apendix \ref{app:3} for more detail),
\begin{align}
\label{eq:D_dis_2}
\mathcal{D}_{xyz,0}^{(\bold{2})}\simeq &c_1\psi_{x,1}(B_y)\psi_{y,0}(B_x)\psi_{z,0}\nonumber\\
&+c_2\psi_{x,2}(B_y)\psi_{y,0}(B_x)\psi_{z,0},
\end{align}
where the perturbative coefficients $c_1$ and $c_2$ are given by,
\begin{align}
    c_1&=\gamma_{1,0}\Bigg[\frac{\langle\psi_{x,0}(B_y)\psi_{z,0}|U_{\mathrm{steps}}|\psi_{x,0}(B_y)\psi_{z,0}\rangle}{\hbar\w_x'}\nonumber\\
    &\qquad-\frac{\langle\psi_{x,1}(B_y)\psi_{z,0}|U_{\mathrm{steps}}|\psi_{x,1}(B_y)\psi_{z,0}\rangle}{\hbar\w_x'}\Bigg],\\
    c_2&=-\gamma_{1,0}\frac{\langle\psi_{x,2}(B_y)\psi_{z,0}|U_{\mathrm{steps}}|\psi_{x,1}(B_y)\psi_{z,0}\rangle}{\hbar\w_x'}.
\end{align}
We now take,
\begin{align}
    \mathcal{D}_{xyz,0}'=\mathcal{D}_{xyz,0}^{(\bold{1})}+\mathcal{D}_{xyz,0}^{(\bold{2})},
\end{align}
that has the same functional form as $\mathcal{D}_{xyz,0}^{(\bold{1})}$ given by \eref{eq:D_dis_1} in which the perturbative coefficients become,
\be\label{eq:gamma'}
  \gamma_{m,n}'=\\
  \begin{cases}
    \gamma_{m,n}+c_1\: , & \{m,n\}=\{1,0\} \, \\
    \gamma_{m,n}+c_2\: , & \{m,n\}=\{2,0\}\, \\
    \gamma_{m,n}\: ,     &  \mathrm{otherwise}\,.
  \end{cases}
\ee
%As we see from \eref{eq:psi_dis}, the effect of disorders is to couple out-of-plane excited states to the in-plane excited states. Since in our model we assumed the miscuts preserve symmetry along $\hat{y}$, \eref{eq:psi_dis} contains in-plane excited sates only along $x$. 

In Figure~\ref{fig:layer_disor}, we  show the electron probability density in the $x-z$ plane in the presence of interface steps. The asymmetry around $x=0$ in this case is due to the change of the quantum-dot thickness due to the interface disorder. Since in our model the Si quantum well is thicker for $x>0$, the peak of the probability density is also shifted towards $x>0$.
\begin{figure}[t!]
\begin{center}
\includegraphics[width=0.45\textwidth]{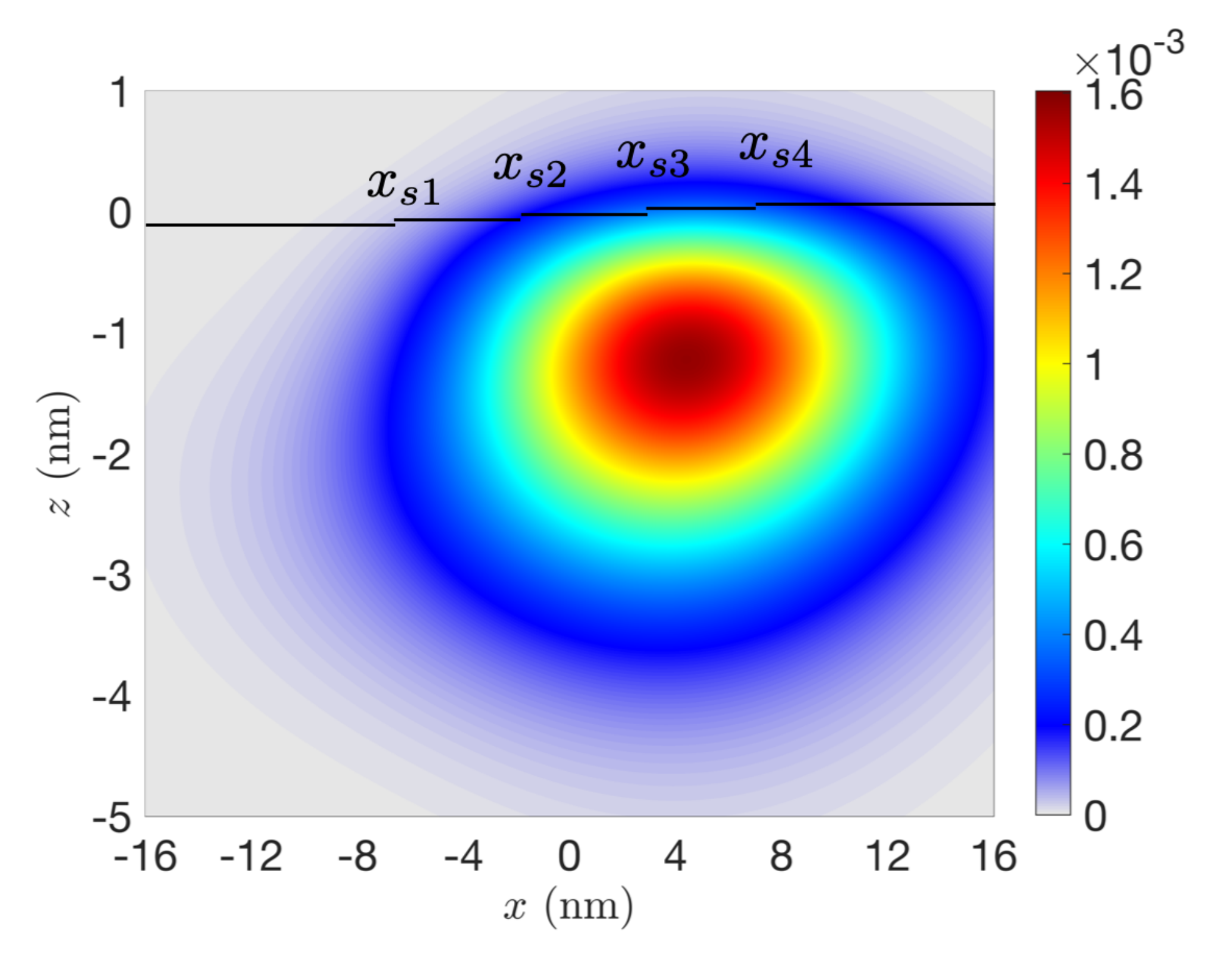}
\end{center}
\caption{The electron probability density in the $x-z$ plane, $|\Psi_{xyz,0}^{\mathrm{dis}}(\bold{B})(x,y=0,z)|^2$ (1/nm$^3$), for a quantum dot with disordered interface. The locations of the interface steps are given in the caption of Table~\ref{tab:param_gamma} and the other parameters are the same as given in the caption of Table~\ref{tab:param}. The solid lines mark the disordered Si/SiGe interface.}  \label{fig:layer_disor}
\end{figure} %======================================

In the next section, we use the envelope functions we found in this section to study and discuss how the valley splitting of a quantum dot depends on the electric and magnetic fields for an ideally flat as well as disordered Si/SiGe interfaces.

\section{Discussion}
\label{sec:Dis}
Within the effective mass theory, the two low-lying valley components of the quantum dot can be written as
\begin{subequations}
\label{eq:valleystates}
\begin{align}
|+z\rangle&=\Psi_{xyz,0}(\textbf{B})e^{ik_0z}u_{+z}(\textbf{r}),\\
|-z\rangle&=\Psi_{xyz,0}(\textbf{B})e^{-ik_0z}u_{-z}(\textbf{r}).
\end{align}
\end{subequations}
Here $k_0\simeq 0.85(2\pi/a_0)$ describes the Bloch wave vector of the conduction band minima and   $u_{\pm z}(r)$ are the periodic parts of the Bloch functions for the $\pm z$ valleys in silicon. We express these functions by a plane wave expansion,
\begin{align}
u_{\pm z}(r)=\sum_\textbf{G} C_{\pm}(\textbf{G})e^{i\textbf{G}.\textbf{r}},
\end{align}
for which $\textbf{G}=(G_x,G_y,G_z)$ is the reciprocal lattice vector. The coefficients in this expansion for the two valleys are related via the time-reversal symmetry relation $C_{-}(\textbf{G})=C_{+}^*(-\textbf{G})$. The wave vectors and their corresponding coefficients $C_{+}(\textbf{G})$ for Si are studied in \ocite{Koiller2011}.

The valley-orbit coupling is given by,
\begin{align}
\Delta_{vo} &=\langle +z\vert-eF_zz +U(z)\vert-z\rangle.
\end{align}
Note that $eF_zz=\e_0\tz$, and given $\e_o\ll U_0$ for all practical values of $F_z$ (see \eref{eq:F_zMax}), the valley-orbit coupling is strongly dominated by the matrix element of the interface potential $U(z)$. Indeed, the role of the electric field is to control and shape $\psi_z$, and the 
contributions from the matrix element of $-eF_zz$ in the valley-orbit coupling can be neglected  \cite{Koiller2011,Koiller2009}.  The valley splitting is found from the above equation by $E_{vs}=2|\Delta_{vo}|$ and the valley phase can be found by 
\begin{align}
\label{eq:vphase}
\phi_\nu=\tan^{-1}\left[\mathrm{Im}(\Delta_{vo})/\mathrm{Re}(\Delta_{vo})\right].
\end{align}

\subsection{Electrical dependence of the valley splitting for an ideal quantum dot}
\label{sec:VS_Fz}
In this section, we consider a quantum dot with an ideal interface in the absence of a magnetic field and use the results of Section \ref{sec:ExEn} to find the electrical and interface-potential dependence of the valley splitting and the valley phase. As depicted in Figure~\ref{fig:Psi_z_vs_Fz}, since the electric field pushes the envelope function towards the upper SiGe barrier, the probability density of the ground state at the lower SiGe/Si interface is negligible. Therefore we take $U(z)=U_0\theta(z)$ and find that the valley-orbit coupling at zero magnetic field for an ideal quantum dot becomes,
\begin{align}
\label{eq:D_vo_0}
\Delta_{vo}^0=U_0\sum_{\bg_1,\bg_2}&\Bigg[C_{+}^*(\bg_1)C_{-}(\bg_2)\nonumber\\
            \times& \int_{-\infty}^{+\infty} e^{-i(\bg_1-\bg_2+2k_0)z}\theta(z)\psi_{z,0}^2dz\Bigg].
\end{align}
To carry on, we note that the terms with $\bg_1\ne \bg_2$ would lead to fast oscillations in the integrand that average to zero. We therefore only consider terms with $\bg_1 = \bg_2$ and define,
\begin{align}
\mathcal{C}_0=\sum_{\bg}C_{+}^*(\bg)C_{-}(\bg),
\end{align}
which we find to be $\mathcal{C}_0=-0.2607$ using \ocite{Koiller2011}.

We now take the integration by parts and find for an ideal quantum dot,
\begin{align}
\label{eq:D_vo_fin}
\Delta_{vo}^0=\Delta_\Int + \Delta_{t}
\end{align}
Here $\Delta_\Int$ is the contribution that comes from the amplitude of $\psi_{z,0}$ at the Si/SiGe interface:
\begin{align}
\label{eq:D_int}
\Delta_\Int = -i\frac{U_0\mathcal{C}_0}{2k_0z_0}\int_{-\infty}^{+\infty} e^{-2ik_0z}\delta(z)\psi_{z,0}^2dz   \:,
\end{align}
and $\Delta_t$ is the contribution that originates from the tail of $\psi_{z,0}$ inside the barrier,
\begin{align}
\Delta_t = -i\frac{U_0\mathcal{C}_0}{2k_0z_0}\int_{-\infty}^{+\infty} e^{-2ik_0z}\theta(z)2\psi_{z,0}\psi_{z,0}'dz   \:.
\end{align}
%============================================
\begin{figure}[b!]
\begin{center}
\includegraphics[width=0.45\textwidth]{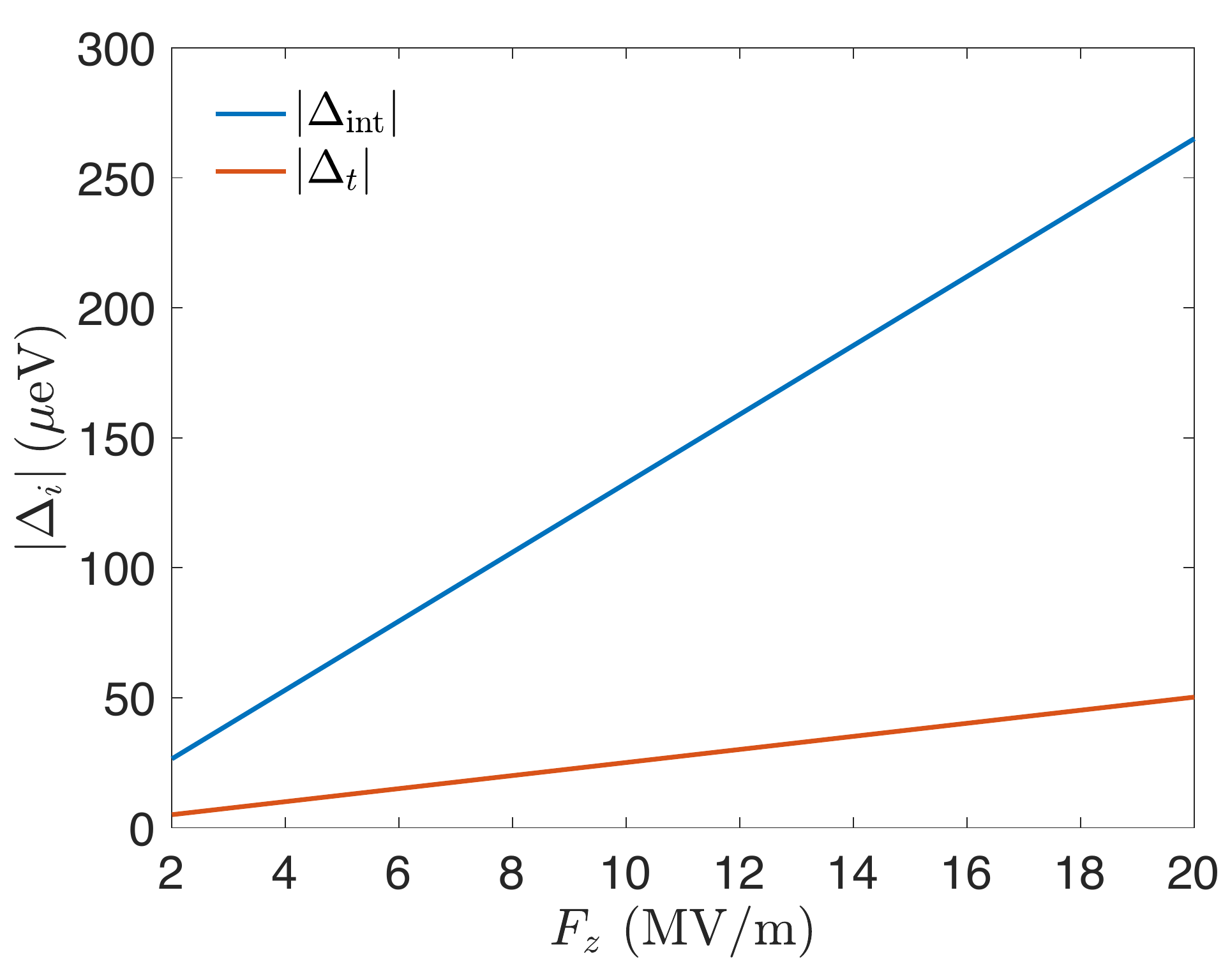}
\end{center}
\caption{$|\Delta_\Int|$ and $|\Delta_t|$ contributions to the valley-orbit coupling for a SiGe barrier. As explained in the main text, the dominant contribution to the valley-orbit coupling comes from the amplitude of the envelope function at the barrier interface described by $|\Delta_\Int|$.}  \label{fig:VO}

\end{figure} %======================================

In order to find analytical expressions for these contributions, we use \esref{eq:E_0z} and (\ref{eq:psi_0z}) and also use the expansions given by \esref{eq:SolExp} (note that by using these equations we assume $U_0\gg \e_0$ which is valid for all relevant valued for the electric field $F_z$, see \eref{eq:F_zMax}.) We finally arrive at the result
\begin{align}
\label{eq:Dint}
\Delta_\Int=-i\mathcal{C}_0\frac{eF_z}{2k_0},
\end{align}
and
\begin{align}
\label{eq:Dt}
\Delta_t=-\Delta_\Int\left[1-\frac{1}{2\tu_0}+i\frac{k_0z_0}{\sqrt{\tu_0}}\right]^{-1}.
\end{align}
The last term in the square bracket is a number larger than $1$ (having $F_z=2$ to $20$ MV/m for a SiGe barrier, we find $k_0z_0/\sqrt{\tu_0} \gtrsim 5$). This indicates that $|\Delta_\Int|$ is larger than $|\Delta_t|$ (by nearly a factor of 6.) 

We conclude that the valley-orbit coupling (and hence also the valley splitting) within the leading order scales linearly with the electric field while it is independent of the interface potential (as long as  $U_0\gg \e_0$). This linear dependence is experimentally observed in \ocite{Yang2013} for a SiO$_2$ barrier (that has a much stronger interface potential $U_0=3$ eV compared with SiGe) and it is also predicted from a theory analysis assuming that the envelope function has zero amplitude inside the barrier \cite{Ruskov2018}.

In addition to the linear-in-electric-field term, here we find the valley-splitting also has a small nonlinear dependence on the electric field (note that $1/\tu_0 \propto F_z^{2/3}$). This non-linear contribution originates from the penetration of the envelope function into the barrier, and can be neglected so long as $\tu_0\gg 1$; this holds provided, 
\begin{align}
\label{eq:F_zMax}
F_z\ll U_0\frac{\sqrt{2m_lU_0}}{e\hbar}.
\end{align}
For a SiGe barrier with $U_0=150$ meV, the right side of the above inequality becomes $\sim 280$ MV/m. This essentially means for all practically relevant values for $F_z$, the valley splitting remains a linear function of the electric field. In Figure~ \ref{fig:VO}, we used \esref{eq:Dint} and (\ref{eq:Dt}) and  show $|\Delta_\Int|$ and $|\Delta_t|$ as a function of the electric field $F_z$. The values that we  show in the figure are in agreement with Refs.~\cite{Culcer2012} and \cite{Koiller2009} where $|\Delta_{vo}|\sim 200\, \mu\mathrm{eV}$ is reported for $F_z=15$ MV/m.

Finally, from \eref{eq:Dint} and (\ref{eq:Dt}) we find that the valley phase of a quantum dot with perfect interface up to the leading order only depends on the interface potential,
\begin{align}
\phi_v\simeq\tan^{-1}\left[\frac{-\hbar k_0}{\sqrt{2m_lU_0}}\right]\:,
\end{align}
that becomes $\phi_v\simeq -0.44\pi$ for a SiGe barrier.

\subsection{Magnetic dependence of the valley splitting for an ideal quantum dot}  
\label{sec:VS_B_clean}
We now extend the results of the last section by including an in-plane magnetic field. As we have shown in Section~\ref{sec:EnvBCle}, the electron envelope function $\Psi_{xyz,0}$ in the presence of an in-plane magnetic field includes excited states of the out-of-plane motion $\psi_{z,n}$, see \eref{eq:Psi_B} and (\ref{eq:Psi_B_1}). Compared to the ground state $\psi_{z,0}$, the excited states can have a larger amplitude at the Si/SiGe interface, and penetrate further to the barrier. As such, we generally expect that the valley splitting should increase in an in-plane magnetic field. In addition, as one can see from Figure \ref{fig:Psi_z_vs_Fz}, depending on the electric field, the excited states can have a sizeable amplitude at the lower SiGe/Si interface $z=-d_t$.  Therefore, we take into account both upper and lower interfaces and consider a barrier potential of the form $U(z)=U_0\theta(z)+U_0\theta(-z-d_t)$. We use \esref{eq:Psi_B} and (\ref{eq:Psi_B_1}) and find for the valley-orbit coupling,
\begin{align}
\label{eq:VS_B}
&\Delta_{vo}^\mathrm{ideal}(\textbf{B},F_z)=\Delta_{vo}^0\left[1+B_x^2B_y^2x_0'^2y_0'^2\eta^2\right]\nonumber \\
&+\mathcal{C}_0B_x^2\left(\frac{y_0'}{z_0}\right)^2\sum_{n,n'}\alpha_n\alpha_{n'}\int_{-\infty}^{\infty}e^{-2ik_0z}\psi_{z,n}U(z)\psi_{z,n'}dz\nonumber\\
&+\mathcal{C}_0B_y^2\left(\frac{x_0'}{z_0}\right)^2\sum_{n,n'}\beta_n\beta_{n'}\int_{-\infty}^{\infty}e^{-2ik_0z}\psi_{z,n}U(z)\psi_{z,n'}dz.
\end{align}
%============================================
\begin{figure}[b!]
\begin{center}
\includegraphics[width=0.45\textwidth]{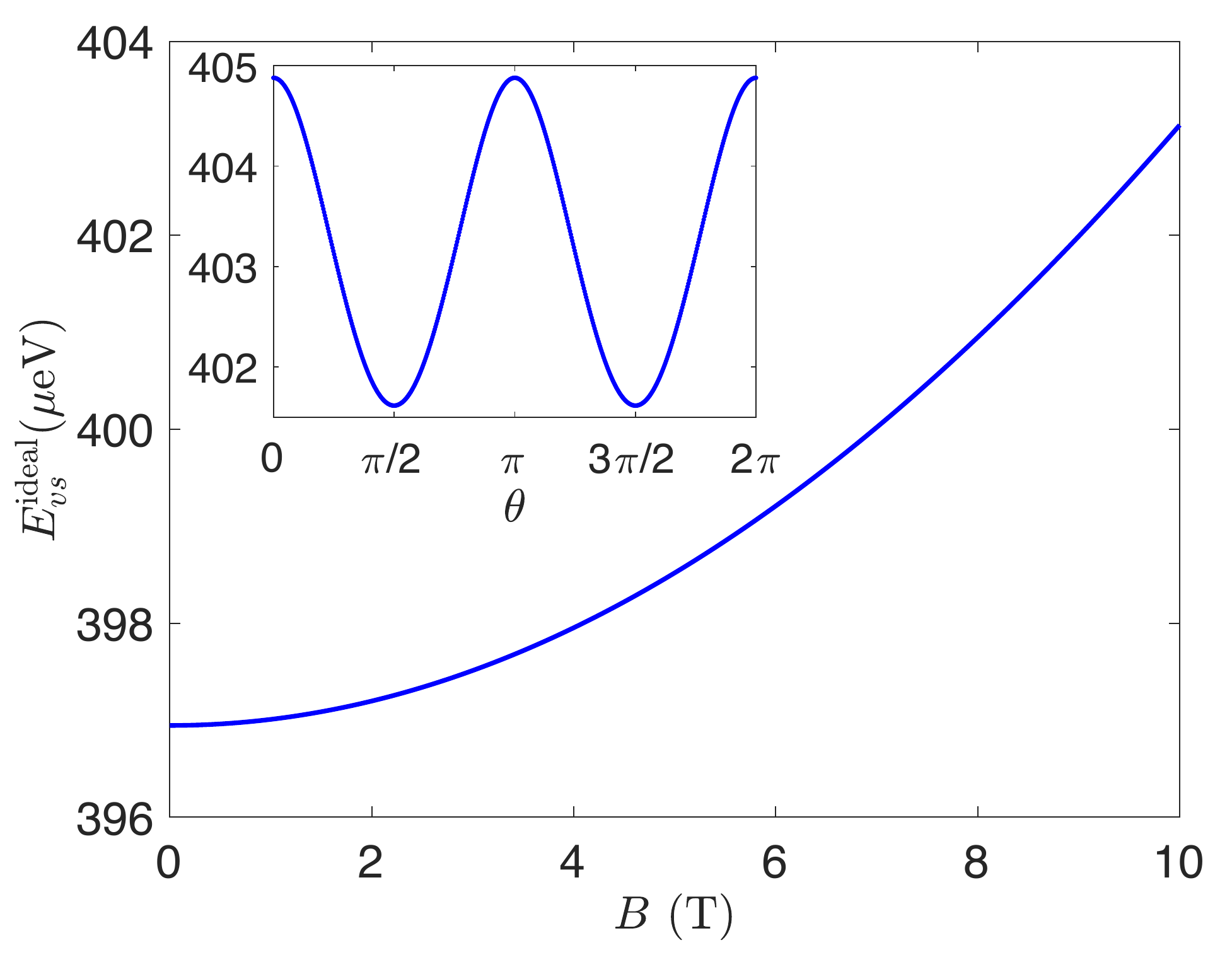}
\end{center}
\caption{Main plot: The valley splitting for a quantum dot with ideal interface at $F_z=15$ MV/m as a function of magnetic field. Here we have taken $\textbf{B}=B(\cos(\theta),\sin(\theta),0)$ and set $\theta=\pi/4$. The size of the quantum dot is the same as indicated in the caption of Table~\ref{tab:param}: $x_0=12$ nm and $y_0=15$ nm. Inset plot: The valley splitting at $B=10$ T as a function of direction of the magnetic field.  }  \label{fig:VS_B_theta}
\end{figure} %======================================

Using the excited states $\psi_{z,n}$ from Section~\ref{sec:ExEn}, we numerically calculate the above integrals. The coefficients $\alpha_n$, $\beta_n$ and $\eta$ are introduced in Section~\ref{sec:EnvBCle} and as we explained there, up to a few Tesla, the terms containing $\alpha$ and $\beta$ are strongly dominant over the correction containing $\eta$. Therefore, to the leading order, the magnetic contribution to the valley splitting scales quadratically with the magnetic field. Moreover, at finite magnetic fields, the valley splitting becomes dependent on ratio of the lateral confinement to the electrical confinement, $x_0'/z_0$ and $y_0'/z_0$- see \esref{eq:x0'} and (\ref{eq:y0'}). 

\eref{eq:VS_B} indicates that for an elliptical quantum dot at a fixed magnetic field, the valley splitting reaches its maximal value when the direction of the magnetic field is perpendicular to the axis with the larger radius. In the main plot of Figure~\ref{fig:VS_B_theta}, we  show the valley splitting as a function of magnetic field at a fixed direction. In the inset plot of the figure, we  show the valley splitting as a function of the direction of the field. We observe that the valley splitting for a quantum dot with ideal interface only slightly increases with the magnetic field. This also indicates that the dominant contribution to the valley splitting remains a linear function of the electric field $F_z$ at the finite magnetic fields.

\subsection{Valley splitting of a quantum dot with disordered interface}  
\label{sec:VS_B_dis}
We now consider a realistic quantum dot with miscuts and steps at the Si/SiGe interface and aim to study the valley splitting and its electromagnetic dependence. We take the interface potential given by \eref{eq:Udis} and use the resulting envelope function \eref{eq:Psi_dis_B}. We then find for the valley-orbit coupling,
\begin{align}
\label{eq:VO_dis_B}
\Delta^\mathrm{dis}_{vo}=N_0^2[&\Delta_{vo}^\mathrm{ideal} + \Delta_{s} + \Delta_{(\textbf{1})} + \Delta_{(\textbf{2})}],
\end{align}
where $\Delta_{vo}^\mathrm{ideal}$ is the valley-orbit coupling for an ideal interface given by \eref{eq:VS_B}, $\Delta_{s}$ is the largest contribution originating from the interface disorders and reads,
\begin{align}
\label{eq:Ds}
&\Delta_{s}=\mathcal{C}_0\int_{-\infty}^{\infty}e^{-2ik_0z}\psi_{x,0}^2(B_y)\psi_{z,0}^2U_\mathrm{steps}dxdz\:,
\end{align}
and $\Delta_{(\textbf{1})}$ is a contribution that is first order with respect to $\mathcal{D}_{xyz,0}'$ and reads,
\begin{align}
\label{eq:D_1}
\Delta_{(\textbf{1})}=\sum_{m,n}\Delta_{(\textbf{1}),\{m,n\}}= 2\mathcal{C}_0\sum_{m,n}\gamma_{m,n}'f_{m,n}\:,
\end{align}
where,
\begin{align}
\label{eq:fmn}
    f_{m,n}=\int_{-\infty}^{\infty}\Big[e^{-2ik_0z}&\psi_{x,m}(B_y)\psi_{z,n}\nonumber\\\times& U_\mathrm{steps}\psi_{x,0}(B_y)\psi_{z,0}\Big]dxdz.
\end{align}
The last term, $\Delta_{(\textbf{2})}$, is a small and sub-leading contribution that is second-order with respect to $\mathcal{D}_{xyz,0}'$ and $\psi_{xyz,0}^{(\bold{1})}$. More details on this term can be found in the Appendix~\ref{app:VS_HOterms}. 

Note that in general we have $|\Delta_{(\textbf{2})}|\ll|\Delta_{(\textbf{1})}|$ and $|\Delta_{(\textbf{1})}|< |\Delta_{vo}|,\: |\Delta_{s}|$. However, as we discuss below, depending on the number and location of interface steps, $|\Delta_{vo} + \Delta_{s}|$ can become a small number.  In this case, the contribution from $\Delta_{(\textbf{1})}$ becomes (more) important in determining the valley splitting and its phase.

\subsubsection{A single step at the interface}
Let us now study the structure and effects of $\Delta_{s}$ and $\Delta_{(\textbf{1})}$. We begin by considering the simplest case; that is, when there is only a single step at the interface.  We take the width of the step to be $+a/4$; the step potential is then obtained from \eref{eq:U_pert} by taking $x_{s4}\rightarrow+\infty$ and $x_{s1}, x_{s2}\rightarrow-\infty$. We then take $x_{s3}=x_s\geq 0$ to be the position of the only interface step. 
From \eref{eq:Ds} for a quantum dot with a single interface step we obtain,
\begin{align} 
\label{eq:D_s_1s}
\Delta_{s}^{1s}(x_s) \simeq &-\frac{1}{2}\Delta_{vo}^0\mathrm{Erfc}\left(\sqrt{2}x_s/x_0'(B_y)\right)\nonumber\\
&\times\left[1-e^{-(a_0/2z_0)(\sqrt{\tu_0}+ik_0z_0)}\right].
\end{align}

In order to analyze $\Delta_{(\textbf{1})}^{1s}$, we note that a complete assessment of this term requires numerical calculations. However, we can obtain a rough estimation by only considering the largest contribution to \eref{eq:D_1}, i.e. the term corresponding to $m=1$ and $n=0$. For further simplicity, we also drop the second-order correction to the envelope function, $\mathcal{D}_{xyz,0}^{(\bold{2})}$, so that we take $\gamma_{1,0}'=\gamma_{1,0}$ and $N_0=1$. We then arrive at the largest contribution to $\Delta_{(\textbf{1})}$ due to $m=1$ and $n=0$,
\begin{align}
\label{eq:D_1_1s}
    \Delta_{(\textbf{1}),\{1,0\}}^{1s}(x_s)\simeq-\gamma_{1,0}&\Delta_{vo}^0\sqrt{\frac{2}{\pi}}e^{-2(x_s/x_0'(B_y))^2}\nonumber\\
    &\times\left[1-e^{-(a_0/2z_0)(\sqrt{\tu_0}+ik_0z_0)}\right],
\end{align}
in which,
\begin{align}
\label{eq:gamma_10}
    \gamma_{1,0}=&\frac{\e_0}{\hbar\omega_x'(B_y)}\frac{1}{2\sqrt{2\pi}}e^{-2(x_s/x_0'(B_y))^2}\nonumber\\
    &\times\frac{1}{\sqrt{\tu_0}}\left[1-e^{-(a/2z_0)\sqrt{\tu_0}}\right].
\end{align}

$\Delta_{s}^{1s}(x_s)$ and $\Delta_{(\textbf{1}),\{1,0\}}^{1s}(x_s)$ are clearly out of phase with $\Delta_{vo}^\mathrm{ideal}$. This can significantly modify and suppress the valley splitting. Note that $\Delta_s^{1s}(x_s)$ and $\Delta_{(\textbf{1}),\{1,0\}}^{1s}(x_s)$ are monotonically decreasing if the interface step is located further away from the quantum-dot center. 

\begin{figure}[t!]
\begin{center}
\includegraphics[width=0.49\textwidth]{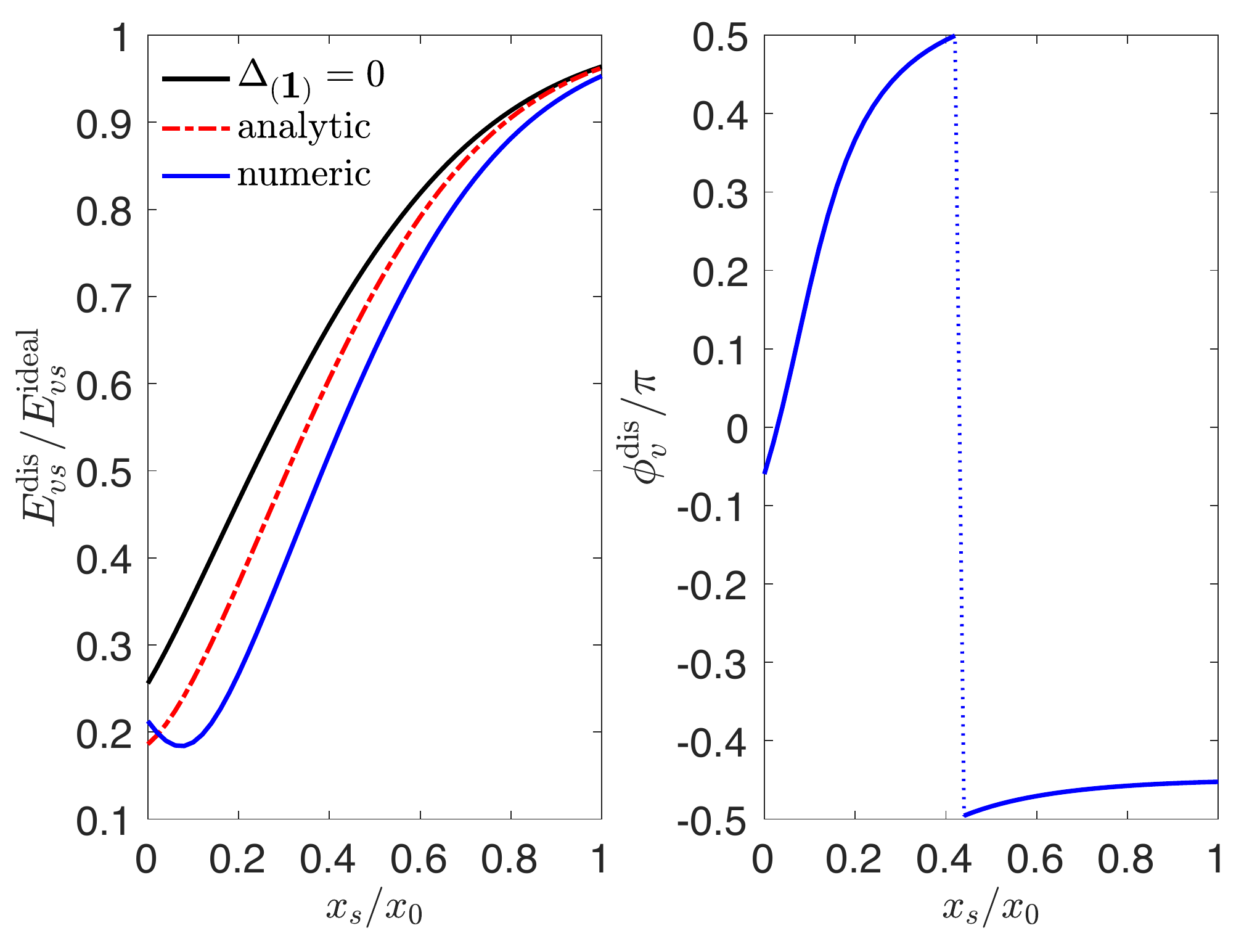}
\end{center}
\caption{Left panel: The normalized valley splitting for a quantum dot with single interface step as a function of the step location. The dashed-dot line is found by using analytic relations \esref{eq:D_vo_fin}, (\ref{eq:D_s_1s}) and (\ref{eq:D_1_1s}) and taking $N_0=1$. The solid lines are deduced from numerical calculation. Right panel:  The valley phase of a quantum dot with a single interface step as a function of the step location. Both panels are obtained at at $B=0$ and $F_z=15$ MV/m. }  \label{fig:1step}
\end{figure}

In the left panel of Figure~\ref{fig:1step} we present the normalized valley splitting as a function of the step location at $B=0$. By neglecting $\Delta_{(\textbf{1})}$ in the valley-orbit coupling (shown in the figure by the solid black line), we observe, as predicted, that the valley splitting is monotonically decreasing as the interface step moves closer to the quantum dot center. At $x=0$, the valley splitting is suppressed by 75\%. This amount of suppression is reported in \ocite{Tariq} where a variational ansatz is used to approximate the envelope function in the presence of a single step. By taking into account the $\Delta_{(\textbf{1}),\{1,0\}}^{1s}(x_s)$ contribution (shown by the dashed-dot red line), we observe that the valley splitting only becomes further suppressed. 

Remarkably, if we numerically calculate $\Delta_{(\textbf{1})}^{1s}$ from \eref{eq:D_1} and take into account not only the dominant term corresponding to $m=1$ and $n=0$ but also the other terms as well, we observe that the valley splitting has in fact a non-monotonic behaviour as a function of the distance between the single interface step and the quantum dot center. In Appendix \ref{app:4}, we discuss in more detail that this behaviour is due to the out-of-plane excited states $\psi_{z,n\geq 1}$.
%============================================
 %======================================%

Indeed, the terms $\Delta_{vo}^\mathrm{ideal}$, $\Delta_s^{1s}$ and $\Delta_{(\textbf{1}),\{1,0\}}^{1s}$ are originating from the ground state of the out-of-plane motion, $\psi_{z,0}$. As it is already shown, the closer the step to the quantum dot center, the further $\Delta_s^{1s}$ and $\Delta_{(\textbf{1}),\{1,0\}}^{1s}$ suppress $\Delta_{vo}^\mathrm{ideal}$, until at some point, $x_s=x_m^{1s}$, the contributions due to the terms with $n\geq 1$ in \eref{eq:D_1} reaches an equal footing as $\Delta_{vo}^\mathrm{ideal}+\Delta_s^{1s}+\Delta_{(\textbf{1}),\{1,0\}}^{1s}$. If the interface step is placed any closer to the quantum dot center, $x_s<x_m^{1s}$, the latter sum continues to decrease whereas the contributions due to the terms with $n\geq 1$ in \eref{eq:D_1} increase due to the increase of the coefficients $\gamma_{m,n\geq 1}$. Therefore, the valley splitting begins to rise. 

In the left panel of the Figure~\ref{fig:1step} we have shown the valley phase as a function of the step location. Note that the valley phase, given by \eref{eq:vphase}, is a $\pi$-periodic function defined within $[-\pi/2,\pi/2]$. Therefore, the sudden jump of the valley phase that occurs at $x_s\sim 0.45 x_0$ can be removed by subtracting $\pi$ from the values above the jump.

Let us now study how the valley splitting scales with the electric and magnetic fields. As we have shown in Sec. \ref{sec:VS_Fz}, the valley-orbit coupling for an ideal quantum dot $\Delta_{vo}^0$ is a linear function of the electric field. As such, the $\Delta_s^{1s}$ term given by \eref{eq:D_s_1s} is also linear in electric field. However,  $\Delta^{1s}_{(\textbf{1})}$ is a non-linear function with respect to the electric field. Given \eref{eq:gamma_10}, the dominant coefficient $\gamma_{1,0}$ scales linearly with the electric field, and this gives rise to a quadratic scaling of $\Delta^{1s}_{(\textbf{1}),\{1,0\}}$ with the electric field (more accurately, if we keep $c_1$ in \eref{eq:gamma'}, it is easy to show that $\Delta^{1s}_{(\textbf{1}),\{1,0\}}$ acquires one more term that is cubic in the electric field.) 

\begin{figure}[t!]
\begin{center}
\includegraphics[width=0.49\textwidth]{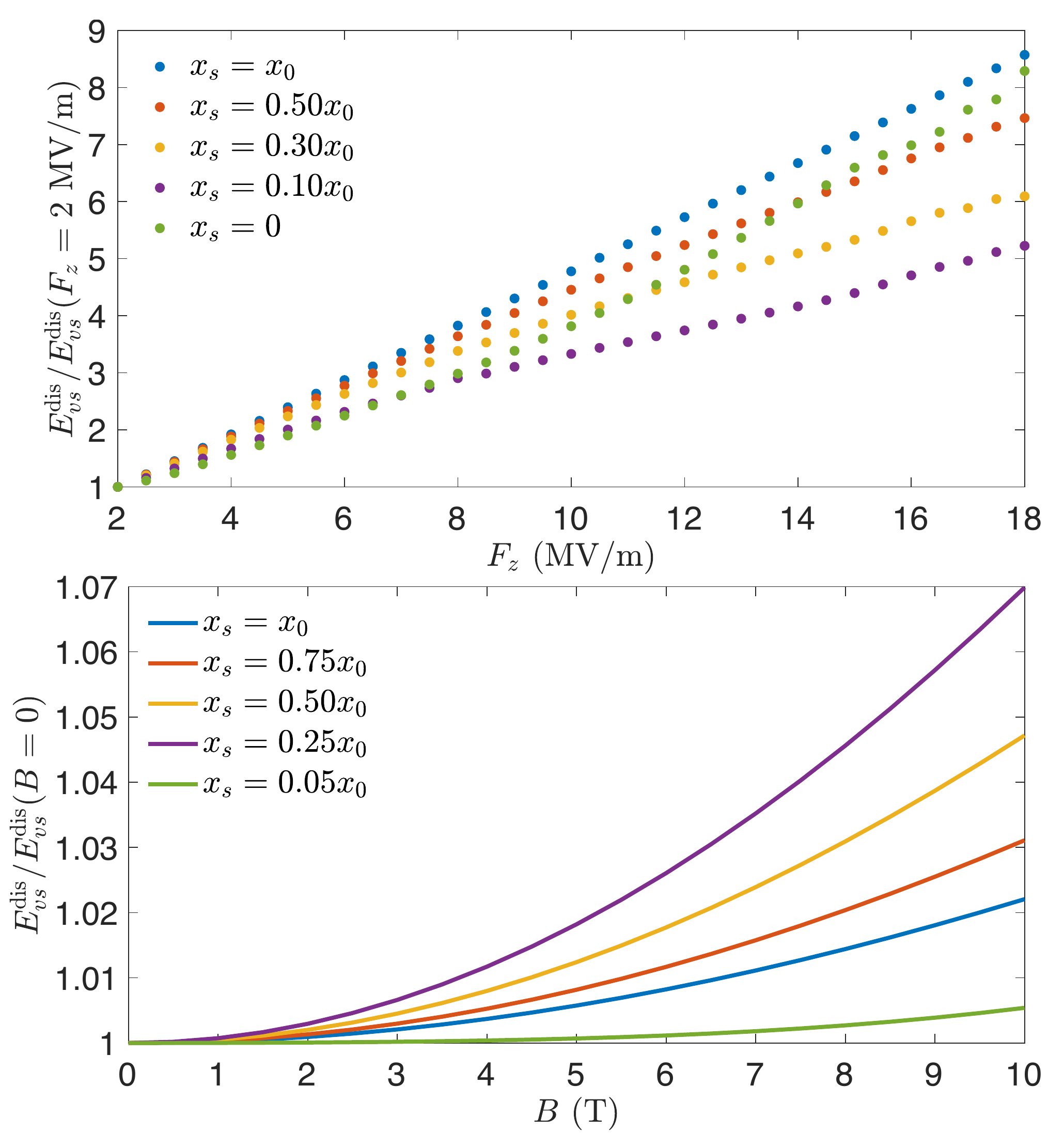}
\end{center}
\caption{Upper panel: The normalized valley splitting for a quantum dot with a single interface step for various step location at $B=0$, as a function of the electric field strength. Lower panel: The normalized valley splitting as a function of the in-plane magnetic field. Here we assumed $\textbf{B}=B(\cos\pi/4, \sin\pi/4, 0)$ and $F_z=15$ MV/m. }  \label{fig:1step_Fz_B}
\end{figure}

Therefore, in general, the valley splitting for a quantum dot with a single interface is a nonlinear function of the electric field due to the $\Delta_{(\textbf{1})}$ term as well as the normalization constant $N_0\simeq [1+\sum_{m,n} \gamma_{m,n}^2]^{-1/2}$. In the upper panel of  Figure~\ref{fig:1step_Fz_B}, we  show the valley splitting as a function of the electric field for several locations for the single interface step. 

Since $\Delta_{(\textbf{1})}$ grows with the electric field faster than a linear function, for $x_s>x_m^{1s}$, we expect the valley splitting should scale sub-linearly by the electric field. The further the step is located away from the center, the smaller $\Delta_{(\textbf{1})}$ becomes so that the valley splitting approaches being a linear function of the electric field. For $x_s<x_m$, the valley splitting is mainly determined from the terms with $n\geq 1$ in $\Delta_{(\textbf{1})}$. Therefore, numerical analysis is required to find the dependency of the valley splitting on the electric field. 

In order to understand how the valley splitting changes with an in-plane magnetic field, we note that the confinement length $x_0'(B_y)$ is reduced by the magnetic field. Using \esref{eq:x0'}, the effect of the magnetic field to the evolution of $\Delta_{s}^{1s}(x_s)$ and $\Delta_{(\textbf{1}),\{1,0\}}^{1s}(x_s)$ is equivalent as if $x'_0=x_0$ and $x_s$ is located from the center at a larger distance, 
\begin{align}
x_s\rightarrow x_s\left(1+\frac{e^2B_y^2}{4\hbar^2}\frac{m_t}{m_l}x_0^4\right)^{1/4}. 
\end{align}
%============================================
 %======================================%

Therefore, we expect that the magnetic field should always increase the valley splitting if $x_s>x_m$. Since the magnetic field effectively increases $x_s$, the increase of the valley splitting by the magnetic field is larger at the step location where the slope of the curve given in the left panel of Figure~\ref{fig:1step} is steeper.  For $x_s<x_m$, the magnetic field decreases $\Delta_{(\textbf{1}),\{m,n\geq 1\}}^{1s}(x_s)$ that controls the valley splitting. However, we observe that the valley splitting still slightly increases as a function of the magnetic field due to the increase of $\Delta^\mathrm{ideal}_{vo}+\Delta_s^{1s}$ with the magnetic field. We display the valley splitting as a function of the in-plane magnetic field in the lower panel of Figure~\ref{fig:1step_Fz_B}.

\subsubsection{Two steps at the interface}
We now consider a quantum dot with two interface steps having widths  $-a_0/4$ and $a_0/4$. The step potential is then obtained from \eref{eq:U_pert} by taking $x_{s1}\rightarrow-\infty$ and $x_{s4}\rightarrow+\infty$. For further simplification, we also assume  that the two steps are placed symmetrically around the center so that we can write $-x_{s2}=x_{s3}=x_s$.  In this case, finding the contribution from $\Delta_{(\textbf{1})}$ requires numerical analysis (even only for the term with $m=1$ and $n=1$ in \eref{eq:D_1}.) However, we can still obtain qualitative understanding of the behaviour of the valley splitting by only considering the effect of $\Delta_s$. In order to arrive to the extension of \eref{eq:Ds} for a quantum dot with two symmetrically located steps, we integrate over $z$ by parts, similar to Sec. \ref{sec:VS_Fz}, and neglect the small integral containing $\psi_{z,0}'$. With this, we arrive at
\begin{align}
\label{eq:D_s_2s}
\Delta_{s}^{2ss}(x_s)&\simeq-\frac{1}{2}\Delta_\mathrm{int}\mathrm{Erfc}\left(\sqrt{2}x_s/x_0'(B_y)\right)\nonumber\\
&\times\Bigg[2-e^{-ik_0a_0/2}\psi_{z,0}(\frac{a_0}{4z_0})^2/\psi_{z,0}(0)^2\nonumber\\
&\qquad-e^{ik_0a_0/2}\psi_{z,0}(-\frac{a_0}{4z_0})^2/\psi_{z,0}(0)^2\Bigg].
\end{align}
%============================================
\begin{figure}[t!]
\begin{center}
\includegraphics[width=0.48\textwidth]{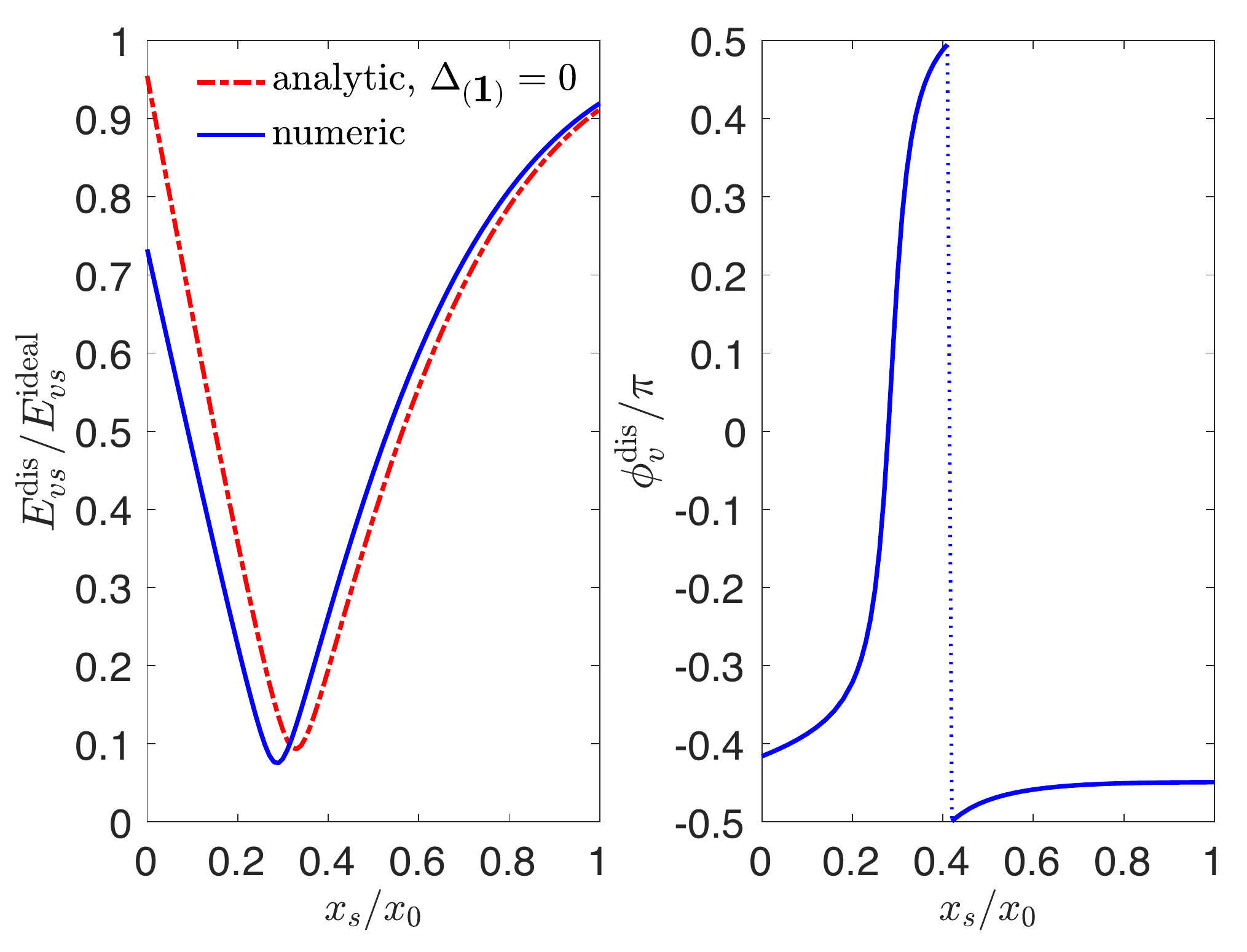}
\end{center}
\caption{Left panel: The normalized valley splitting for a quantum dot with two interface steps.  Right Panel:  The valley phase as a function of the steps' location. In both panels we used $B=0$ and $F_z=15$ MV/m.}  \label{fig:VS_vs_2ss}
\end{figure} %======================================

If the steps are located at the center, $x_s=0$, we find from the above equation (taking $F_z=15$ MV/m)  $\Delta_{s}^{2ss}(x_s=0)\simeq (1.93-0.18i)\Delta_\mathrm{int}$ that is larger than $\Delta_{vo}^\mathrm{ideal}$ in \eref{eq:VO_dis_B} (note that $\Delta_{vo}^\mathrm{ideal}$ is largely determined by $\Delta_\mathrm{int}$; see \esref{eq:D_vo_fin} and (\ref{eq:VS_B}) and Fig.~\ref{fig:VO}.) When the steps are located away from the center, it reduces $\Delta_{s}^{2ss}(x_s)$ so that eventually at some point $|\Delta_{vo}^\mathrm{ideal}|$ becomes larger than $|\Delta_{s}^{2ss}|$. As such, we can expect that the valley splitting of a quantum dot with two symmetrically located steps is a also non-monotonic function  of the steps' location, $x_s$. 

This behaviour is clearly shown in the left panel of Figure~\ref{fig:VS_vs_2ss}. The dashed-dot line of the figure shows the valley splitting obtained by neglecting $\Delta_{(\textbf{1})}$, using \esref{eq:D_vo_fin}, (\ref{eq:VS_B}) and (\ref{eq:D_s_2s}), and setting $N_0=1$.  The blue line is obtained by numerical calculation with all terms in \eref{eq:VO_dis_B}  included. We observe that the effect of $\Delta_{(\textbf{1})}$ is to further suppress the valley splitting as well as shift the step location where the valley splitting reaches its minimum. At this step location, $x_s=x_m^{2ss}\sim 0.3x_0$, the valley splitting is suppressed by more than $90\%$. The right panel of the figure shows how the valley phase is changed as a function of the step location. As mentioned before, the sudden jump can be removed by using the $\pi$-periodicity of the valley phase.

\begin{figure}[t!]
\begin{center}
\includegraphics[width=0.50\textwidth]{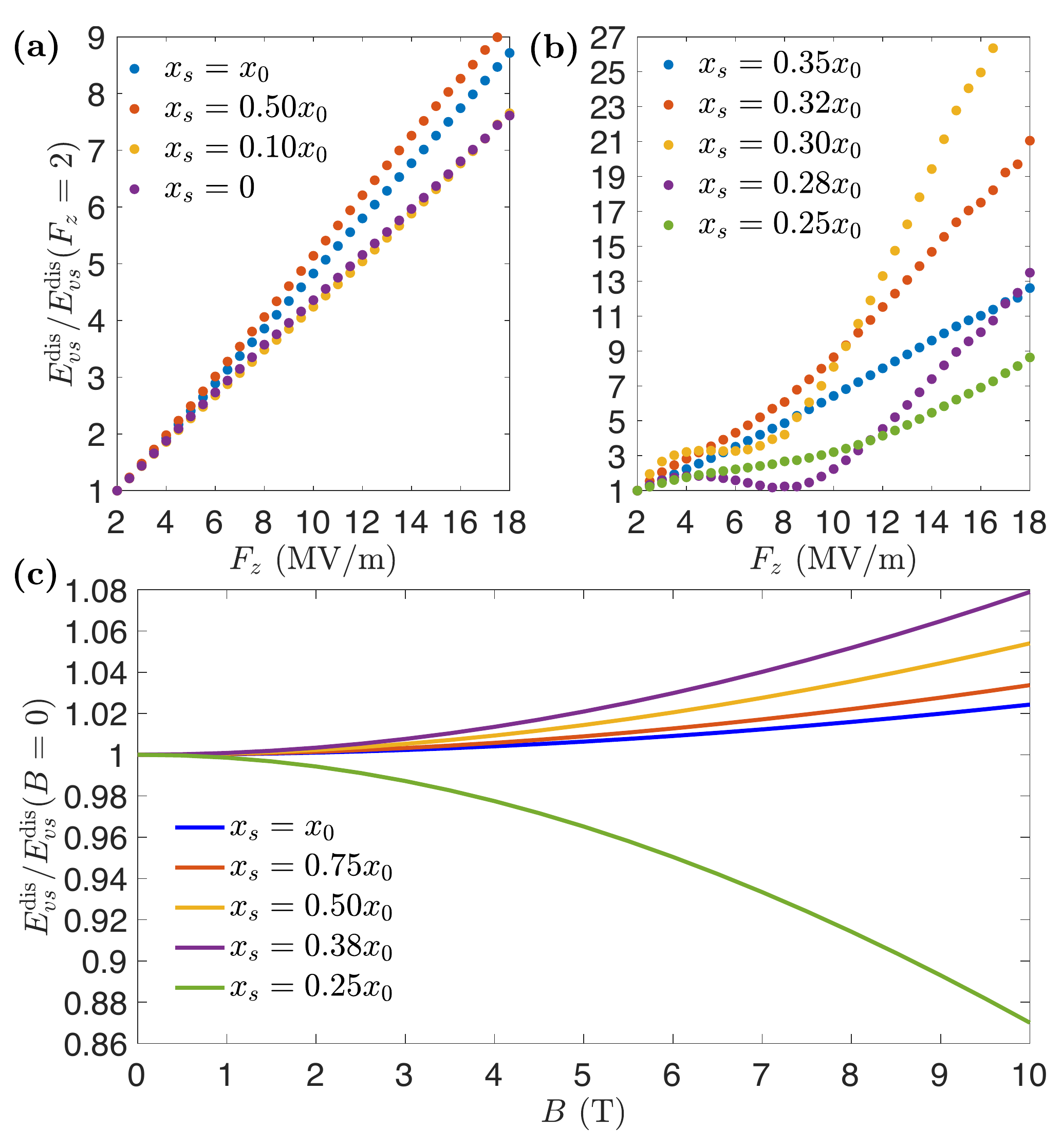}
\end{center}
\caption{\textbf{(a)} and \textbf{(b)}:  Normalized valley splitting as a function of the step location and electric field at $B=0$. \textbf{(c)}: Normalized valley splitting as a function of an in-plane magnetic field. Similar to the lower panel of Fig. \ref{fig:1step_Fz_B}, here we assumed $\textbf{B}=B(\cos\pi/4, \sin\pi/4, 0)$ and $F_z=15$ MV/m.}  \label{fig:VS_vs_Fz_B_2ss}
\end{figure}

We now study the electromagnetic dependence of the valley splitting for a quantum dot with two symmetrically locates interface steps. Given Fig. \ref{fig:VS_vs_2ss}, away from $x_m^{2ss}$, the dominant contribution to the valley splitting is due to $\Delta_{vo}^\mathrm{ideal}$ and $\Delta_s$. As such, we expect the scaling of the valley splitting with the electric field has to be approximately linear. This behaviour is shown in the panel (a) of the Figure~\ref{fig:VS_vs_Fz_B_2ss}. At the step locations close to $x_m^{2ss}$, the contribution due to $\Delta_{(\textbf{1})}$ becomes important so that we can expect a strong non-linear dependency on the electric field; this is shown in the panel (b) of the Figure.

In the panel (c) of the figure we  show how the valley splitting is changed by the magnetic field. For $x_s<x_m^{2ss}$, the valley splitting is mainly controlled by $\Delta_{s}^{2ss}$. As such, since this term reduces by the magnetic field, the valley splitting is also decreasing as well. For $x_s>x_m^{2ss}$, $\Delta_{vo}^\mathrm{ideal}$ is the dominant contribution to the valley splitting; therefore, the magnetic field always increases the valley splitting by reducing $\Delta_{s}^{2ss}$ (and $\Delta_{(\textbf{1})}$.) Note that this effect is stronger for the step locations where the slope of the curve shown in left panel of Fig. \ref{fig:VS_vs_2ss} is steeper.
%============================================
 %===============================
%============================================
\begin{figure}[b!]
\begin{center}
\includegraphics[width=0.45\textwidth]{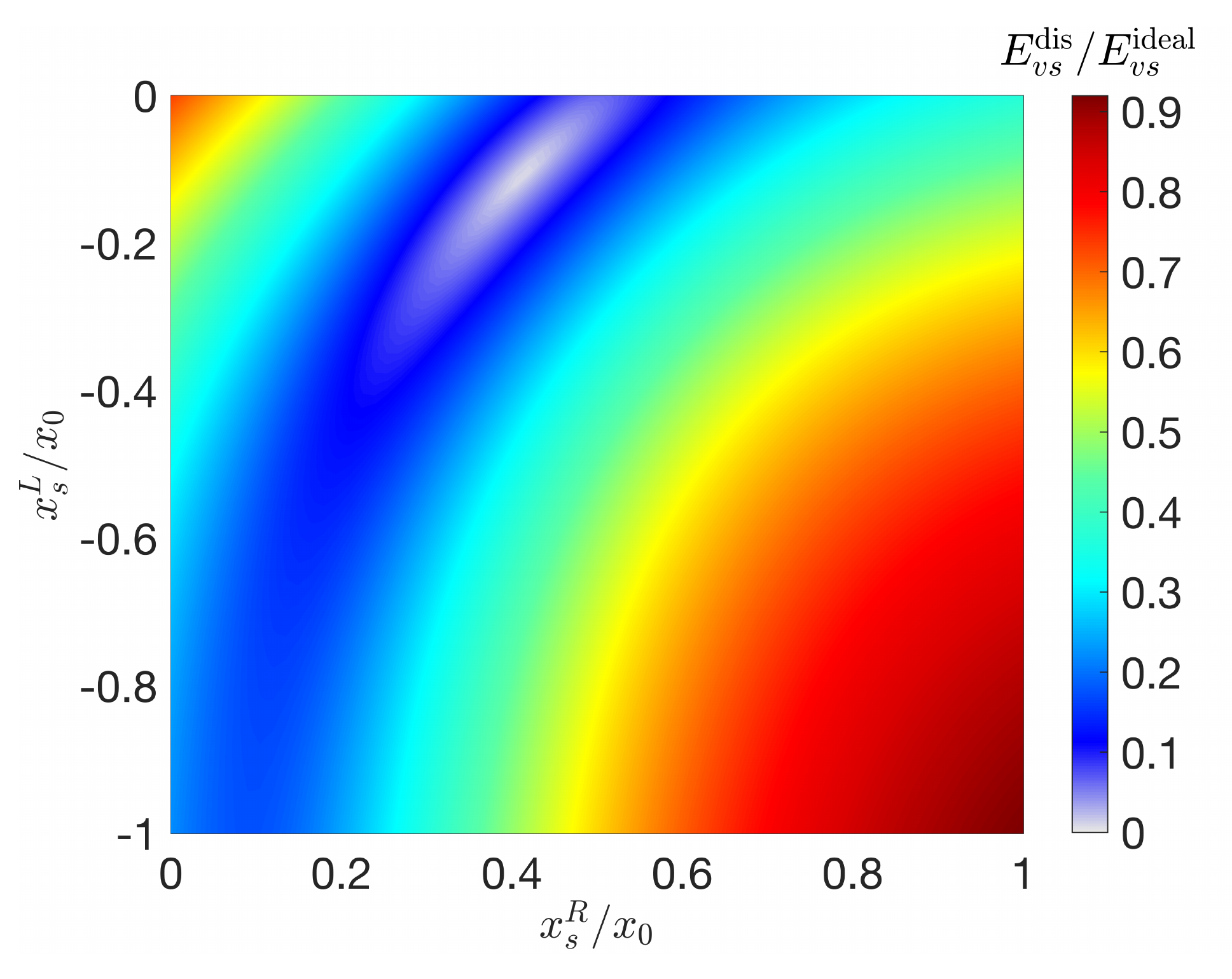}
\end{center}
\caption{The normalized valley splitting as a function of position of two interface steps at $B=0$ and $F_z=15$ MV/m.}  \label{fig:VS_2s}
\end{figure} %===============================

Finally, we relax the condition of the two steps being symmetric around the quantum-dot center and use the step positions $x_{s2}=x_s^L\leq 0$ and $x_{s3}=x_s^R\geq 0$ in \eref{eq:U_pert}.  In Figure~\ref{fig:VS_2s}, we present our result for the normalized valley splitting as a function of the position of each step. Note that due to the non-symmetric nature of the step potential, the valley splitting turns out not to be symmetric with respect to the location of the steps. We observe that the valley splitting can  completely vanish in some specific configuration of the interface steps; this is also predicted in \ocite{Tariq} using a simpler model.
We have also studied the valley splitting for models including three and four interface steps, and we observed qualitatively similar behavior for the valley splitting as a function of electric and magnetic fields as presented in this section.

\section{Summary and conclusions}
To summarize, the valley splitting is one of the important properties for the silicon quantum dots that directly influences the lifetime and scalability of  silicon spin qubits. As such, understanding the behaviour and tunability of the valley splitting is very important. In this work, we studied how the valley splitting responds to the electromagnetic field for both ideal and disordered quantum dots. We considered a realistic potential profile for a SiGe/Si/SiGe quantum dot by taking into account both lower and upper Si/SiGe interfaces as well the interface between upper SiGe layer and the insulating layer hosting the gate electrodes; see Fig. \ref{fig:qubit_layout} and \eref{eq:Uz}. While so far the out-of-plane electron motion has been studied by variational methods in a simpler potential model including only one Si/SiGe interface, we found the exact (within effective mass theory) envelope functions of the ground state as well as the excited states for the out-of-plane electron motion. This has enabled us to find the electron envelope function in the presence of finite magnetic field as well as interface disorder. In both cases, the envelope function reflects the coupling between in-plane to out-of-plane degrees of freedom, see \esref{eq:Psi_B} and (\ref{eq:Psi_dis_B}). Our analysis enables us to obtain the coupling coefficients using  perturbation theory, for arbitrary configurations for the interface disorder.  

We showed that in an ideal quantum dot, the valley splitting, within the leading order, always scales linearly with the out-of-plane electric field; see Fig.~\ref{fig:VO}. Moreover, the valley splitting slightly increases with an applied in-plane magnetic field due to the coupling to the out-of-plane excited states; see Figure~\ref{fig:VS_B_theta}. The presence of interface disorder can significantly modify and suppress the valley splitting. We considered a stair-like disordered interface and studied the suppression of valley splitting due to the interface miscuts.

For a quantum dot with a disordered interface, we found that depending on the number and locations of the interface steps, the valley splitting can scale non-linearly with the electric field; see the upper panel of Figure \ref{fig:1step_Fz_B} and panels (a) and (b) of Figure~\ref{fig:VS_vs_Fz_B_2ss}. If there is only one miscut at the interface, the magnetic field always increases the valley splitting, see the lower panel of Figure~\ref{fig:1step_Fz_B}. However, for multiple interface miscuts, the magnetic field can both increase or even suppress the valley splitting, depending on the  configuration of the miscuts, see panel (c) of Figure~\ref{fig:VS_vs_Fz_B_2ss}.

In the theory of spin relaxation induced by the valley coupling, one important set of quantities are the  transition dipole matrix elements between the valley states \cite{Yang2013,Huang14}. For an ideal quantum dot, the envelope function has an in-plane mirror symmetry (Fig.~\ref{fig:2D_psi}). This, in turn, gives rise to vanishing of the in-plane dipole matrix elements. However, the presence of the interface disorder can break the in-plane mirror symmetry (Fig.~\ref{fig:layer_disor}). Our findings for the envelope function in the presence of disorder  now enable the prediction of the dipole matrix elements as a function of the electromagnetic field. To our knowledge, the dipole matrix elements have always been taken as fitting parameters. Future work will be needed to further develop the theory of spin relaxation induced by the valley coupling based on a calculation of the transition dipole matrix elements and valley splitting in a magnetic field. 
\label{sec:conc}

\acknowledgments
We gratefully acknowledge useful discussions with Maximilian Russ and Mónica Benito. This work has been supported by ARO grant number W911NF-15-1-0149.

\appendix
\section{Ground state of a triangular potential well: A self-consistent approximation}
\label{app:GS_SC}
In this appendix, we present our analysis leading to the ground state energy and wavefunction given by \esref{eq:E_0z} and (\ref{eq:psi_0z}). We first make the reasonable assumption that the ground state energy is much smaller than the interface potential $E_{z,0}\ll U_0$. This assumption enables us to neglect the interface between the upper SiGe barrier with the insulating layer at $z=d_t$. Moreover, since the electric field pushes the envelope function towards the upper SiGe barrier, we also neglect the lower SiGe interface at $z=-d_t$ as the probability amplitude in that region is very small.  Therefore, we simplify the full interface potential given by \eref{eq:Uz} for the ground state and take it as $U(z)=U_0\theta(z)$.

We can then use the confinement length and energy given by \esref{eq:z0}-(\ref{eq:e0}) to define the  dimensionless quantity,
\be\label{eq:zeta0}
  \tze_0(z)=\\
  \begin{cases}
\tilde{U}_0-\tz-\te_{z,0}\: , & z > 0 \, \\
-\tz-\te_{z,0}\:                 . &  z \leq  0 \, 
  \end{cases}
\ee
We then arrive to the below Schr\"odinger equation for the envelope function 
\begin{align}
\label{eq:H_z0}
\frac{d^2}{d\tz^2}\psi_{z,0}-\tze_{0}\psi_{z,0}=0.
\end{align}
Inside the silicon layer where $z\leq0$, $\psi_{z,0}$ is given by the Airy function of the first kind,
\begin{subequations}
\label{eq:psi_z_an}
\begin{align}
\label{eq:psi_z_l0}
\psi_{z,0} = M_0z_0^{-1/2}\Ai(-\tz-\tez),\quad z\leq 0.
\end{align}

Let us now consider the form of the eigenstate inside the SiGe barrier;  the exact solution for the envelope function, up to prefactors, reads Ai($\tilde{U}_0-\tz-\te_{z,0}$). However, in order to find an analytic relation for the ground state energy, we try to approximate the envelope function in the barrier. If there was no electric field inside the barrier (i.e. the $-\tilde{z}$ term in  $\tze_0(z>0)$ was absent), the eigenstates would have been proportional to exp($-\sqrt{\tu_0-\tez}z$). Due to the presence of the electric field, the potential barrier is reduced along $z$ and the wavefunction can further penetrate into the barrier. To take this into account, we introduce a parameter $\lambda$ into the exponent of the exponentially decaying wavefunction that allows further penetration into SiGe provided $\lambda<1$. We thus approximate the wavefunction inside the barrier by,
\begin{align}
\label{eq:psi_z_g0}
\psi_{z,0} \simeq M_0z_0'^{-1/2}\Ai(-\tez)e^{-\lambda\sqrt{\tu_0-\tez}z}.\quad z> 0
\end{align}
\end{subequations}

Using the continuity of the first derivative of $\psi_{z,0}$ at the $z=0$ interface, we can write the penetration parameter $\lambda$ as,
\begin{align}
\label{eq:gamma}
\lambda=\frac{\Ai'(-\tez)}{\Ai(-\tez)}\frac{1}{\sqrt{\tu_0-\tez}},
\end{align}
We now self-consistently determine the ground state energy by noting,
\begin{align}
\tez=\frac{\langle \psi_{z,0}\vert \frac{d^2}{d\tz^2}-\tz+\tu_0\theta(\tz) \vert\psi_{z,0}\rangle}{\langle \psi_{z,0} \vert\psi_{z,0}\rangle}.
\end{align}
From the above equation and using \esref{eq:psi_z_an} and (\ref{eq:gamma}) we finally arrive at,
\begin{align}
\label{eq:Efin}
(\tilde{U}_0-\tez)&\Ai'(-\tez)\Ai^2(-\tez)\nonumber\\
&=\frac{1}{2}\Ai^3(-\tez)+\Ai'^3(-\tez).
\end{align}
In order to find the solution of the \eref{eq:Efin}, we note that for the infinite potential well $\tilde{U}_0=\infty$, the ground state energy is determined by the smallest root (in absolute value)  of the Airy function, $-r_0\simeq -2.3381$. This suggests that for a finite but high potential well $\tilde{U}_0\gg 1$, the solution should remain close to $-r_0$.  As such, we consider a solution of the form,
\begin{align}
\label{eq:E_expan}
\tez = r_0 + \delta\te,
\end{align}
and expand Ai and Ai$'$  functions around $-r_0$. We  keep terms up to quadratic order in $\delta\te$ to find,
\begin{subequations}
\label{eq:SolExp}
\begin{align}
&\Ai(-\tez)=-\delta\te\Ai'(-r_0) + \mathcal{O}(\delta\te^3),\\
&\Ai'(-\tez)=\Ai'(-r_0)-\frac{1}{2}\delta\te^2r_0\Ai'(-r_0) + \mathcal{O}(\delta\te^3).
\end{align}
\end{subequations}
By substituting \esref{eq:SolExp} into \eref{eq:Efin} and keeping terms up to quadratic order we find $\tilde{U}_0\delta\te^2=1$ that gives $\delta\te=\pm 1/\sqrt{\tilde{U}_0}$. 

In order to determine which sign is physically acceptable, we note from \esref{eq:psi_z_g0} and (\ref{eq:gamma}) that the wave function decays inside the barrier provided $\Ai'(-\tez)/\Ai(-\tez)>0$. This is satisfied only if $\delta\te$ has a negative sign. In this case, we also find $\lambda<1$, as expected.  Note that if we have kept the expansions in \eref{eq:SolExp} up to the cubic order, we would have found a correction to the $\tez$ of order $\tu_0^{-3/2}$.

Finally, from the normalization of the envelope function we find,
\begin{align}
\label{eq:M0} 
M_0= \Big[\tez&\Ai(-\tez)^2+\Ai'(-\tez)^2 \nonumber\\
&+\frac{1}{2}\Ai(-\tez)^3/\Ai'(-\tez)\Big]^{-1/2}\:.
\end{align}
In order to simplify this, we use \esref{eq:E_expan} and (\ref{eq:SolExp}) and find $M_0\simeq 1/\Ai'(-r_0)$.
\section{Valley splitting of a disordered quantum dot in magnetic field: Higher-order terms}
\label{app:VS_HOterms}
In this Appendix, we present the form of the sub-leading contribution $\Delta_{(\textbf{2})}$ in \eref{eq:VO_dis_B}. 
Using \esref{eq:Udis} and  (\ref{eq:Psi_dis_B}) we find 
\begin{widetext}
\begin{align}
\Delta^{(\textbf{2})}(\textbf{B})=\mathcal{C}_0\Bigg\{&B_x^2\left(\frac{y_0'}{z_0}\right)^2\sum_{n,n'}\alpha_n\alpha_{n'}\int_{-\infty}^{\infty}e^{-2ik_0z}\psi_{x,0}(B_y)^2\psi_{z,n}\psi_{z,n'}U_\mathrm{steps}(x,z)dxdz\nonumber\\
+&B_y^2\left(\frac{x_0'}{z_0}\right)^2\sum_{n,n'}\beta_n\beta_{n'}\int_{-\infty}^{\infty}e^{-2ik_0z}\psi_{x,1}(B_y)^2\psi_{z,n}\psi_{z,n'}U_\mathrm{steps}(x,z)dxdz\nonumber\\
+&B_x^2B_y^2x_0'^2y_0'^2\eta^2\int_{-\infty}^{\infty}e^{-2ik_0z}\psi_{x,1}(B_y)^2\psi_{z,0}^2U_\mathrm{steps}(x,z)dxd\nonumber\\+&\sum_{m,n,n'=1}\gamma_{m,n}\gamma_{m,n'}\int_{-\infty}^{\infty}U(z)e^{-2ik_0z}\psi_{z,n}\psi_{z,n'}dz\nonumber\\
+&\sum_{m,n,m',n'}\gamma_{m,n}\gamma_{m',n'}\int_{-\infty}^{\infty}e^{-2ik_0z}\psi_{z,n}\psi_{z,n'}\psi_{x,m}(B_y)\psi_{x,m'}(B_y)U_\mathrm{steps}(x,z)dxdz\Bigg\}\:.
\end{align}
\end{widetext}
The perturbative coefficients $\alpha$, $\beta$, $\eta$ and $\gamma$ are given in Sections \ref{sec:EnvBCle} and \ref{sec:EnvBDis}, and using the excited states $\psi_{z,n}$ from Section \ref{sec:ExEn}, we can numerically evaluate the integrals.

\section{Second-order correction to the envelope function due to the interface disorder}
\label{app:3}
Here we present the complete form of the second order correction due to the interface steps. According to  perturbation theory, we have
\begin{align}
    \mathcal{D}_{xyz,0}^{(\textbf{2})}=&\sum_{\{m',n'\}\ne\{m,n\}}\sum_{\{m,n\}\ne\{0,0\}}\gamma_{m,n}\zeta_{m',n'}^{m,n}\psi_{x,m'}\psi_{z,n'}\nonumber\\
    &-\sum_{\{m,n\}\ne\{0,0\}} \gamma_{m,n}c_{m,n}\psi_{x,m'}\psi_{z,n'},
\end{align}
where we defined,
\begin{align}
    &\zeta_{m',n'}^{m,n}=\frac{\langle\psi_{x,m'}\psi_{z,n'}|U_{\mathrm{steps}}|\psi_{x,m}\psi_{z,n}\rangle}{E_{n,z}-E_{n',z}+(m-m')\hbar\w_x'},\\
   &c_{m,n}=\frac{\langle\psi_{x,0}\psi_{z,0}|U_{\mathrm{steps}}|\psi_{x,0}\psi_{z,0}\rangle}{E_{0,z}-E_{n,z}-m\hbar\w_x'}.
\end{align}
In order to arrive to \eref{eq:D_dis_2}, we only keep the dominant terms; i.e., in the set of $\gamma_{m,n}$, we keep $\gamma_{1,0}$, and in the set of $\zeta_{m',n'}^{m,n}$, we keep $\zeta_{2,0}^{1,0}$ and $\zeta_{0,0}^{1,0}$. Finally, in the set of $c_{m,n}$, we keep $c_{1,0}$.

\section{The effect of out-of-plane excited states in $\Delta_{(\textbf{1})}$ }\label{app:4}
%============================================
\begin{figure}[b]
\begin{center}
\includegraphics[width=0.49\textwidth]{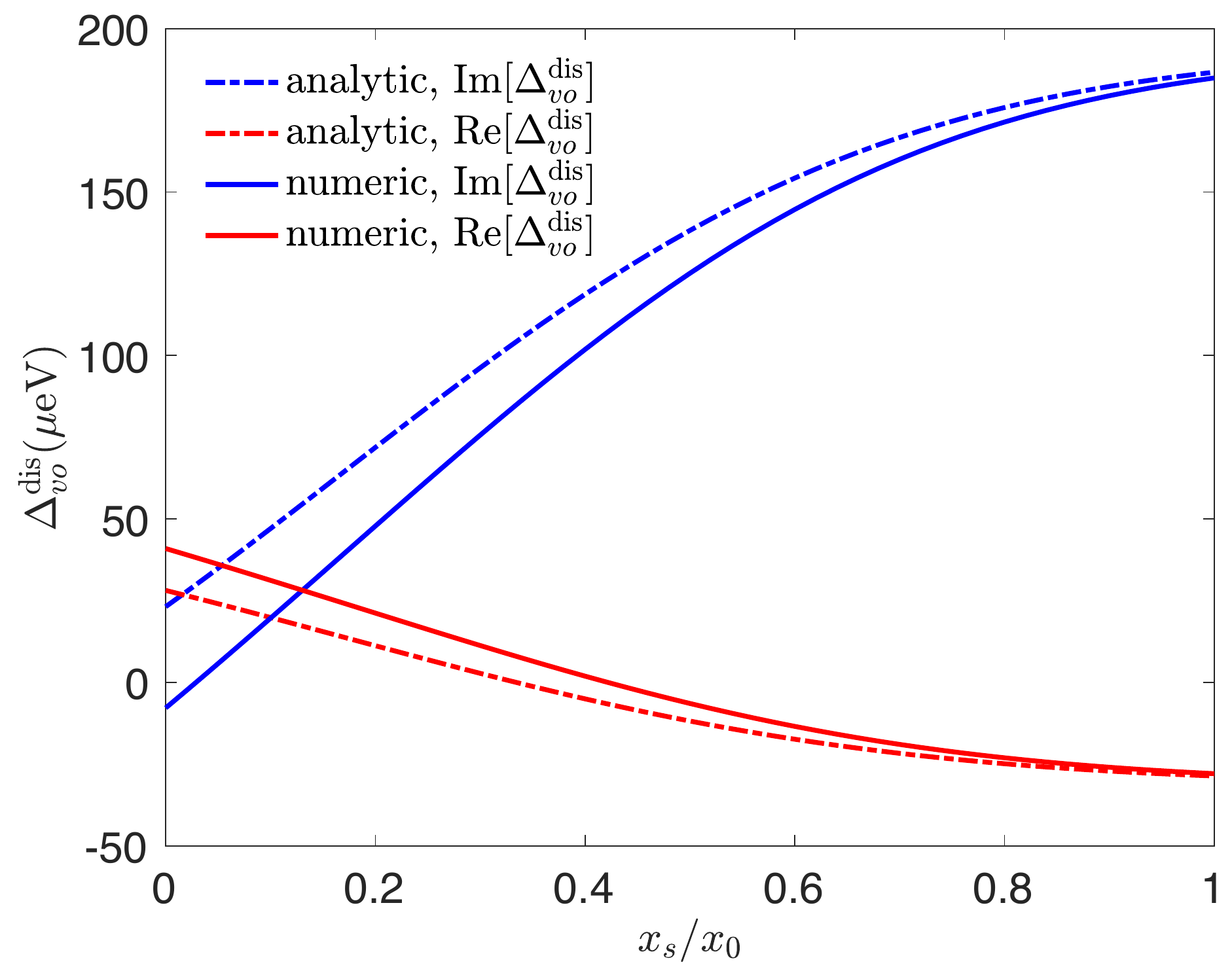}
\end{center}
\caption{Imaginary and real parts of the valley-orbit coupling for a quantum dot with single interface step at $\textbf{B}=0$ and $F_z=15$ MV/m. In order to obtain the dashed-dot lines, we used $\Delta_{vo}^\mathrm{ideal}$ from \eref{eq:D_vo_fin}, $\Delta_{s}^{1s}$ from \eref{eq:D_s_1s} and $\Delta_{(\textbf{1}),\{1,0\}}^{1s}$ from \eref{eq:D_1_1s}. For the solid lines, we numerically calculated all contributions in \eref{eq:VO_dis_B} including terms with $n\geq 1$ in $\Delta_{(\textbf{1}),\{m,n\}}^{1s}$. }  \label{fig:Re_Im_VO}
\end{figure} %======================================%
In this appendix, we explain in more detail the effect of the excited states $\psi_{z,n\geq 1}$ to the valley splitting for a quantum dot with single interface step. The step potential \eref{eq:U_pert} in this case becomes,
\begin{align}
    U_{\mathrm{steps}}^{1s}(x,z)=-U_0\theta(x-x_s)\theta(z)\theta\left(z-\frac{a_0}{4}\right)
\end{align}
In Section~\ref{sec:VS_B_dis}, we found the contributions from  $\Delta_{s}^{1s}$ and $\Delta_{(\textbf{1}),\{1,0\}}^{1s}$ are out-of-phase with $\Delta_{vo}^\mathrm{ideal}$, therefore,  these terms would monotonically suppress the valley splitting. Here we show that once the out-of-plane excited states $\psi_{z,n}$ are taken into account, the real part of $\Delta_{(\textbf{1})}^{1s}$  gives rise to a non-monotonic behaviour of the valley splitting as a function of the step location, as shown in left panel of Fig. \ref{fig:1step}.

To see this effect, let us integrate \eref{eq:fmn} over z by parts to find,
\begin{align}\label{eq:f_mn_appendix}
    &f_{m,n}\simeq iU_0\frac{1}{2k_0}\int_{x_s}^{\infty}\psi_{x,m}\psi_{x,0}dx\nonumber\\
    &\times\left[\psi_{z,n}(0)\psi_{
    z,0}(0)-e^{-ik_0a_0/2}\psi_{z,n}(\frac{a_0}{4})\psi_{
    z,0}(\frac{a_0}{4})\right],
\end{align}
where we have neglected the small contributions containing integral over $e^{-2ik_0z}\psi_{z,n}\psi_{z,0}'$ and $e^{-2ik_0z}\psi_{z,n}'\psi_{z,0}$. As we have shown in Sec.~\ref{sec:VS_Fz}, the dominant contribution in $\Delta_{vo}^\mathrm{ideal}$ is given by $\Delta_\mathrm{int}$ from \eref{eq:Dint} which is an imaginary quantity and is due to the amplitude of the envelope function at the Si/SiGe interface.

Given \eref{eq:f_mn_appendix} and \eref{eq:D_1}, the imaginary part of $\Delta_{(\textbf{1})}^{1s}$ is in opposite phase with $\Delta_\mathrm{int}$ (note that in \eref{eq:f_mn_appendix} we have $e^{-ik_0a_0/2}=-0.891 - 0.454i$.) Therefore, the closer the step is located to the quantum dot center, the more the imaginary part of total valley-orbit coupling is suppressed. On the other hand, the real part of the valley-orbit coupling is increased when the step is closer to the center, and at some point it becomes the dominant contribution to the total valley-orbit coupling. We plot the imaginary and real parts of the valley orbit coupling in Fig.~\ref{fig:Re_Im_VO}. The dashed-dot lines are obtained by using the analytic relations we obtained in Sec.~\ref{sec:VS_B_dis} for $D_s^{1s}$ and $D_{(\textbf{1}),\{1,0\}}^{1s}$ whereas the solid lines are found from numerically evaluating the valley-orbit coupling including all terms in $\Delta_{(\textbf{1}),\{m,n\}}^{1s}$. We observe that the non-monotonic behaviour of the valley splitting as a function of step location can be seen only by taking into account the out-of-plane excited states $\psi_{z,n}$.

%=============================================================================================


\begin{thebibliography}{30}
\bibitem{Zwanenburg13} F. A. Zwanenburg, A. S. Dzurak, A. Morello, M. Y. Simmons, L. C. L. Hollenberg, G. Klimeck, S. Rogge, S. N. Coppersmith, and M. A. Eriksson, \href{https://doi.org/10.1103/RevModPhys.85.961}{Rev. Mod. Phys. \textbf{85}, 961 (2013).}

\bibitem{Morello10} A. Morello, J. J. Pla, F. A. Zwanenburg, K. W. Chan, K. Y. Tan, H. Huebl, M. Möttönen, C. D. Nugroho, C. Yang, J. A. van Donkelaar, A. D. C. Alves, D. N. Jamieson, C. C. Escott, L. C. L. Hollenberg, R. G. Clark, and A. S. Dzurak, \href{https://doi.org/10.1038/nature09392}{Nature  \textbf{467}, 687 (2010).}


\bibitem{Yang2013} C. H. Yang, A. Rossi, R. Ruskov, N. S. Lai, F. A. Mohiyaddin, S. Lee, C. Tahan, G. Klimeck, A. Morello and A. S. Dzurak, \href{https://doi.org/10.1038/ncomms3069}{Nat Commun \textbf{4}, 2069 (2013).}

\bibitem{Borjans19} F. Borjans, D.M. Zajac, T.M. Hazard, and J.R. Petta, \href{https://doi.org/10.1103/PhysRevB.83.165301}{Phys. Rev. Applied \textbf{11}, 044063 (2019).}

\bibitem{Assali11} L. V. C. Assali, H. M. Petrilli, R. B. Capaz, B. Koiller, X. Hu, and S. Das Sarma, \href{https://doi.org/10.1103/PhysRevB.83.165301}{Phys. Rev. B \textbf{83}, 165301 (2011).}

\bibitem{Tyryshkin12} A. M. Tyryshkin, S. Tojo, J. J. L. Morton, H. Riemann, N. V. Abrosimov, P. Becker, H.-J. Pohl, T. Schenkel, M. L. W. Thewalt, K. M. Itoh, and S. A. Lyon, \href{https://doi.org/10.1038/nmat3182}{Nature Mater \textbf{11}, 143–147 (2012)}

\bibitem{Steger12} M. Steger, K. Saeedi, M. L. W. Thewalt, J. J. L. Morton, H. Riemann, N. V. Abrosimov, P. Becker, and H.-J. Pohl, \href{https://10.1126/science.1217635}{Science \textbf{336}, 1280 (2012).}

\bibitem{Mi18} X. Mi, M. Benito, S. Putz, D. M. Zajac, J. M. Taylor, G. Burkard and J. R. Petta, \href{https://doi.org/10.1038/nature25769}{Nature \textbf{555}, 599–603 (2018)}

\bibitem{Samkharadze18} N. Samkharadze, G. Zheng, N. Kalhor, D. Brousse, A. Sammak, U. C. Mendes, A. Blais, G. Scappucci and L. M. K. Vandersypen, \href{doi.org//10.1126/science.aar4054}{Science \textbf{359}, 1123–1127 (2018).}

\bibitem{Veldhorst14} M. Veldhorst, J. C. C. Hwang, C. H. Yang, A. W. Leenstra, B. de Ronde, J. P. Dehollain, J. T. Muhonen, F. E. Hudson, K. M. Itoh, A. Morello and A. S. Dzurak, \href{https://doi.org/10.1038/nnano.2014.216}{Nature Nanotech \textbf{9}, 981–985 (2014).}

\bibitem{Yoneda18} J. Yoneda, K. Takeda, T. Otsuka, T. Nakajima, M. R. Delbecq, G. Allison, T. Honda, T. Kodera, S. Oda, Y. Hoshi, N. Usami, K. M. Itoh and S. Tarucha, \href{https://doi.org/10.1038/s41565-017-0014-x}{Nature Nanotech \textbf{13}, 102–106 (2018).}

\bibitem{Zajac18}
	D. M. Zajac, A. J. Sigillito, M. Russ, F. Borjans,  J. M. Taylor, G. Burkard, and J. R. Petta,
	\href{https://science.sciencemag.org/content/359/6374/439}
	{Science {\bf 359}, 439--442 (2018).}


\bibitem{Watson18} T. F. Watson, S. G. J. Philips, E. Kawakami, D. R. Ward, P. Scarlino, M. Veldhorst, D. E. Savage, M. G. Lagally, M. Friesen, S. N. Coppersmith, M. A. Eriksson and L. M. K. Vandersypen, \href{https://doi.org/10.1038/nature25766}{Nature \textbf{555}, 633–637 (2018).}

\bibitem{Dzurak19_tqg} W. Huang, C. H. Yang, K. W. Chan, T. Tanttu, B. Hensen, R. C. C. Leon, M. A. Fogarty, J. C. C. Hwang, F. E. Hudson, K. M. Itoh, A. Morello, A. Laucht and A. S. Dzurak \href{https://doi.org/10.1038/s41586-019-1197-0}{Nature \textbf{569}, 532–536 (2019).}

\bibitem{Golub04}L. E. Golub and E. L. Ivchenko, \href{https://doi.org/10.1103/PhysRevB.69.115333}{Phys. Rev. B 69, 115333 (2004).}

\bibitem{Veldhorst15R} M. Veldhorst, R. Ruskov, C. H. Yang, J. C. C. Hwang, F. E. Hudson, M. E. Flatté, C. Tahan, K. M. Itoh, A. Morello, and A. S. Dzurak, \href{https://doi.org/10.1103/PhysRevB.92.201401}{Phys. Rev. B \textbf{92}, 201401(R) (2015).}

\bibitem{Ferdous18} R. Ferdous, E. Kawakami, P. Scarlino, M. P. Nowak, D. R. Ward, D. E. Savage, M. G. Lagally, S. N. Coppersmith, M. Friesen, M. A. Eriksson, L. M. K. Vandersypen and R. Rahman, \href{https://doi.org/10.1038/s41534-018-0075-1}{\textit{npj Quantum Inf} \textbf{4}, 26 (2018).}

\bibitem{Huang14} P. Huang and X. Hu, \href{https://doi.org/10.1103/PhysRevB.90.235315}{Phys. Rev. B \textbf{90}, 235315 (2014).}

\bibitem{Friesen07} M. Friesen, S. Chutia, C. Tahan, and S. N. Coppersmith,\href{https://doi.org/10.1103/PhysRevB.75.115318}{Phys. Rev. B 75, 115318 (2007).} 

\bibitem{Koiller2011} A. L. Saraiva, M. J. Calderón, Rodrigo B. Capaz, Xuedong Hu, S. Das Sarma, and Belita Koiller, \href{https://doi.org/10.1103/PhysRevB.84.155320}{Phys. Rev. B \textbf{84}, 155320 (2011).}

\bibitem{Culcer2012} Y. Wu and D. Culcer, \href{https://doi.org/10.1103/PhysRevB.86.035321}{Phys. Rev. B \textbf{86}, 035321 (2012).}

\bibitem{Tariq} B. Tariq and X. Hu, \href{https://doi.org/10.1103/PhysRevB.100.125309}{Phys. Rev. B \textbf{100}, 125309 (2019).}

\bibitem{Jock2018} R. M. Jock, N. T. Jacobson,P. Harvey-Collard, \textit{et al.}, \href{https://doi.org/10.1038/s41467-018-04200-0}{Nat Commun \textbf{9}, 1768 (2018).}

\bibitem{Zandvliet93}H. J. W. Zandvliet and H. B. Elswijk,\href{https://doi.org/10.1103/PhysRevB.48.14269}{Phys. Rev. B \textbf{48}, 14269 (1993).}

\bibitem{Hollmann19} A. Hollmann, T. Struck, V. Langrock, A. Schmidbauer, F. Schauer, K. Sawano, H. Riemann, N. V. Abrosimov, D. Bougeard, L. R. Schreiber, \href{https://arxiv.org/abs/1907.04146}{arXiv:1907.04146}

\bibitem{Friesen06} M. Friesen, M. A. Eriksson, and S. N. Coppersmith\href{https://doi.org/10.1063/1.2387975},
{Appl. Phys. Lett. \textbf{89}, 202106 (2006).}

\bibitem{Goswami06} S. Goswami, K. A. Slinker, M. Friesen, L. M. McGuire, J. L. Truitt, C. Tahan, L. J. Klein, J. O. Chu, P. M. Mooney, D. W. van der Weide, R. Joynt, S. N. Coppersmith, M. A. Eriksson, \href{https://doi.org/10.1038/nphys475}{Nature Phys \textbf{3}, 41–45 (2007).}

\bibitem{Davies} See, e.g., \href{https://doi.org/10.1017/CBO9780511819070}{ \textit{The physics of low-dimensional semiconductors}} by  J. H Davies. Cambridge Univercity Press (1998).

\bibitem{Koiller2009} A. L. Saraiva, M. J. Calderón, X. Hu, S. Das Sarma, and B. Koiller, \href{https://doi.org/10.1103/PhysRevB.80.081305}{Phys. Rev. B \textbf{80}, 081305(R) (2009)}


\bibitem{Ruskov2018} R. Ruskov, M. Veldhorst, A. S. Dzurak, and C. Tahan, \href{https://doi.org/10.1103/PhysRevB.98.245424}{Phys. Rev. B \textbf{98}, 245424 (2018).}

\end{thebibliography}
\end{document}